\documentclass[trackchanges, twocolumn]{aastex7}
\shorttitle{The DESI Transients Survey}
\shortauthors{Hall, Palmese, et al.}

\newcommand{\mcwilliams}{
    McWilliams Center for Cosmology and Astrophysics,
    Department of Physics,
    Carnegie Mellon University,
    5000 Forbes Avenue, Pittsburgh, PA 15213
}

\begin{document}

\title{The DESI Transients Survey: Legacy Classifications and Methodology}



\author[0000-0002-9364-5419]{Xander J. Hall}
\affiliation{\mcwilliams}
\email[show]{xhall@cmu.edu}

\author[0000-0002-6011-0530]{Antonella Palmese}
\affiliation{\mcwilliams}
\email{palmese@cmu.edu}

\author[0000-0001-5537-4710]{Segev BenZvi}
\affiliation{Department of Physics \& Astronomy, University of Rochester, 206 Bausch and Lomb Hall, P.O. Box 270171, Rochester, NY 14627-0171, USA}
\email{sbenzvi@ur.rochester.edu}

\author[0000-0003-0776-8859]{John Banovetz}
\affiliation{Lawrence Berkeley National Laboratory, 1 Cyclotron Road, Berkeley, CA 94720, USA}
\email{jdbanovetz@lbl.gov}

\author[0000-0002-9700-0036]{Brendan O'Connor}
\affiliation{\mcwilliams}
\email{boconno2@andrew.cmu.edu}

\author[0000-0001-7201-1938]{Lei Hu}
\affiliation{\mcwilliams}
\email{leihu@andrew.cmu.edu}

\author[0000-0002-5698-8703]{Erica Hammerstein}
\affiliation{Department of Astronomy, University of California, Berkeley, CA 94720-3411, USA}
\affiliation{Lawrence Berkeley National Laboratory, 1 Cyclotron Road, Berkeley, CA 94720, USA}
\email{ekhammer@berkeley.edu}

\author[0000-0003-3433-2698]{Ariel Amsellem}
\affiliation{\mcwilliams}
\email{aamselle@andrew.cmu.edu}

\author{Jessica Nicole Aguilar}
\affiliation{Lawrence Berkeley National Laboratory, 1 Cyclotron Road, Berkeley, CA 94720, USA}
\email{jaguilar@lbl.gov}

\author[0000-0001-6098-7247]{Steven Ahlen}
\affiliation{Department of Physics, Boston University, 590 Commonwealth Avenue, Boston, MA 02215 USA}
\email{ahlen@bu.edu}

\author[0000-0003-4162-6619]{Steven Bailey}
\affiliation{Lawrence Berkeley National Laboratory, 1 Cyclotron Road, Berkeley, CA 94720, USA}
\email{stephenbailey@lbl.gov}

\author[0000-0001-9712-0006]{Davide Bianchi}
\affiliation{Dipartimento di Fisica ``Aldo Pontremoli'', Universit\`a degli Studi di Milano, Via Celoria 16, I-20133 Milano, Italy}
\affiliation{INAF-Osservatorio Astronomico di Brera, Via Brera 28, 20122 Milano, Italy}
\email{davide.bianchi1@unimi.it}

\author{David Brooks}
\affiliation{Department of Physics \& Astronomy, University College London, Gower Street, London, WC1E 6BT, UK}
\email{david.brooks@ucl.ac.uk}

\author{Todd Claybaugh}
\affiliation{Lawrence Berkeley National Laboratory, 1 Cyclotron Road, Berkeley, CA 94720, USA}
\email{tmclaybaugh@lbl.gov}

\author[0000-0002-2169-0595]{Andrei Cuceu}
\affiliation{Lawrence Berkeley National Laboratory, 1 Cyclotron Road, Berkeley, CA 94720, USA}
\email{acuceu@lbl.gov}

\author[0000-0002-0553-3805]{Kyle Dawson}
\affiliation{Department of Physics and Astronomy, The University of Utah, 115 South 1400 East, Salt Lake City, UT 84112, USA}
\email{kdawson@astro.utah.edu}

\author[0000-0002-1769-1640]{Axel de la Macorra}
\affiliation{Instituto de F\'{\i}sica, Universidad Nacional Aut\'{o}noma de M\'{e}xico,  Circuito de la Investigaci\'{o}n Cient\'{\i}fica, Ciudad Universitaria, Cd. de M\'{e}xico  C.P. 04510,  M\'{e}xico}
\email{macorra@fisica.unam.mx}

\author[0000-0003-0928-2000]{John Della Costa}
\affiliation{Department of Astronomy, San Diego State University, 5500 Campanile Drive, San Diego, CA 92182, USA}
\affiliation{NSF NOIRLab, 950 N. Cherry Ave., Tucson, AZ 85719, USA}
\email{john.dellacosta@noirlab.edu}

\author[0000-0002-4928-4003]{Arjun Dey}
\affiliation{NSF NOIRLab, 950 N. Cherry Ave., Tucson, AZ 85719, USA}
\email{arjun.dey@noirlab.edu}

\author{Peter Doel}
\affiliation{Department of Physics \& Astronomy, University College London, Gower Street, London, WC1E 6BT, UK}
\email{apd@star.ucl.ac.uk}

\author[0000-0003-4992-7854]{Simone Ferraro}
\affiliation{Lawrence Berkeley National Laboratory, 1 Cyclotron Road, Berkeley, CA 94720, USA}
\affiliation{University of California, Berkeley, 110 Sproul Hall \#5800 Berkeley, CA 94720, USA}
\email{sferraro@lbl.gov}

\author[0000-0002-3033-7312]{Andreu Font-Ribera}
\affiliation{Institut de F\'{i}sica d'Altes Energies (IFAE), The Barcelona Institute of Science and Technology, Edifici Cn, Campus UAB, 08193, Bellaterra (Barcelona), Spain}
\email{afont@ifae.es}

\author[0000-0002-2890-3725]{Jaime E. Forero-Romero}
\affiliation{Departamento de F\'isica, Universidad de los Andes, Cra. 1 No. 18A-10, Edificio Ip, CP 111711, Bogot\'a, Colombia}
\affiliation{Observatorio Astron\'omico, Universidad de los Andes, Cra. 1 No. 18A-10, Edificio H, CP 111711 Bogot\'a, Colombia}
\email{je.forero@uniandes.edu.co}

\author[0000-0001-9632-0815]{Enrique Gaztañaga}
\affiliation{Institute of Cosmology and Gravitation, University of Portsmouth, Dennis Sciama Building, Portsmouth, PO1 3FX, UK}
\affiliation{Institut d'Estudis Espacials de Catalunya (IEEC), c/ Esteve Terradas 1, Edifici RDIT, Campus PMT-UPC, 08860 Castelldefels, Spain}
\email{gaztanaga@gmail.com}

\author[0000-0003-3142-233X]{Satya Gontcho A Gontcho}
\affiliation{Lawrence Berkeley National Laboratory, 1 Cyclotron Road, Berkeley, CA 94720, USA}
\affiliation{University of Virginia, Department of Astronomy, Charlottesville, VA 22904, USA}
\email{satya@virginia.edu}

\author[0000-0003-4089-6924]{Alma Xochitl Gonzalez-Morales}
\affiliation{Departamento de F\'{\i}sica, DCI-Campus Le\'{o}n, Universidad de Guanajuato, Loma del Bosque 103, Le\'{o}n, Guanajuato C.~P.~37150, M\'{e}xico}
\email{gonzalez.alma@ugto.mx}

\author[0000-0002-4391-6137]{Or Graur}
\affiliation{Institute of Cosmology and Gravitation, University of Portsmouth, Dennis Sciama Building, Portsmouth, PO1 3FX, UK}
\email{or.graur@port.ac.uk}

\author{Gaston Gutierrez}
\affiliation{Fermi National Accelerator Laboratory, PO Box 500, Batavia, IL 60510, USA}
\email{gaston@fnal.gov}

\author[0000-0002-6024-466X]{Mustapha Ishak}
\affiliation{Department of Physics, The University of Texas at Dallas, 800 W. Campbell Rd., Richardson, TX 75080, USA}
\email{mishak@utdallas.edu}

\author[0000-0001-8528-3473]{Jorge Jimenez}
\affiliation{Institut de F\'{i}sica d’Altes Energies (IFAE), The Barcelona Institute of Science and Technology, Edifici Cn, Campus UAB, 08193, Bellaterra (Barcelona), Spain}
\email{jjimenez@ifae.es}

\author[0000-0003-0201-5241]{Dick Joyce}
\affiliation{NSF NOIRLab, 950 N. Cherry Ave., Tucson, AZ 85719, USA}
\email{richard.joyce@noirlab.edu}

\author[0000-0002-0000-2394]{Stephanie Juneau}
\affiliation{NSF NOIRLab, 950 N. Cherry Ave., Tucson, AZ 85719, USA}
\email{stephanie.juneau@noirlab.edu}

\author[0000-0001-6356-7424]{Anthony Kremin}
\affiliation{Lawrence Berkeley National Laboratory, 1 Cyclotron Road, Berkeley, CA 94720, USA}
\email{akremin@lbl.gov}

\author{Ofer Lahav}
\affiliation{Department of Physics \& Astronomy, University College London, Gower Street, London, WC1E 6BT, UK}
\email{o.lahav@ucl.ac.uk}

\author[0000-0002-6731-9329]{Claire Lamman}
\affiliation{The Ohio State University, Columbus, 43210 OH, USA}
\email{lamman.1@osu.edu}

\author[0000-0003-1838-8528]{Martin Landriau}
\affiliation{Lawrence Berkeley National Laboratory, 1 Cyclotron Road, Berkeley, CA 94720, USA}
\email{mlandriau@lbl.gov}

\author[0000-0001-7178-8868]{Laurent Le Guillou}
\affiliation{Sorbonne Universit\'{e}, CNRS/IN2P3, Laboratoire de Physique Nucl\'{e}aire et de Hautes Energies (LPNHE), FR-75005 Paris, France}
\email{llg@lpnhe.in2p3.fr}

\author[0000-0002-3677-3617]{Alexie Leauthaud}
\affiliation{Department of Astronomy and Astrophysics, UCO/Lick Observatory, University of California, 1156 High Street, Santa Cruz, CA 95064, USA}
\affiliation{Department of Astronomy and Astrophysics, University of California, Santa Cruz, 1156 High Street, Santa Cruz, CA 95065, USA}
\email{alexie@ucsc.edu}

\author[0000-0003-1887-1018]{Michael Levi}
\affiliation{Lawrence Berkeley National Laboratory, 1 Cyclotron Road, Berkeley, CA 94720, USA}
\email{melevi@lbl.gov}

\author[0000-0003-4962-8934]{Marc Manera}
\affiliation{Departament de F\'{i}sica, Serra H\'{u}nter, Universitat Aut\`{o}noma de Barcelona, 08193 Bellaterra (Barcelona), Spain}
\affiliation{Institut de F\'{i}sica d'Altes Energies (IFAE), The Barcelona Institute of Science and Technology, Edifici Cn, Campus UAB, 08193, Bellaterra (Barcelona), Spain}
\email{mmanera@ifae.es}

\author[0000-0002-1125-7384]{Aaron Meisner}
\affiliation{NSF NOIRLab, 950 N. Cherry Ave., Tucson, AZ 85719, USA}
\email{aaron.meisner@noirlab.edu}

\author{Ramon Miquel}
\affiliation{Instituci\'{o} Catalana de Recerca i Estudis Avan\c{c}ats, Passeig de Llu\'{\i}s Companys, 23, 08010 Barcelona, Spain}
\affiliation{Institut de F\'{i}sica d'Altes Energies (IFAE), The Barcelona Institute of Science and Technology, Edifici Cn, Campus UAB, 08193, Bellaterra (Barcelona), Spain}
\email{rmiquel@ifae.es}

\author[0000-0002-2733-4559]{John Moustakas}
\affiliation{Department of Physics and Astronomy, Siena University, 515 Loudon Road, Loudonville, NY 12211, USA}
\email{jmoustakas@siena.edu}

\author{Adam Myers}
\affiliation{Department of Physics \& Astronomy, University  of Wyoming, 1000 E. University, Dept.\textasciitilde{}3905, Laramie, WY 82071, USA}
\email{amyers14@uwyo.edu}

\author[0000-0001-9070-3102]{Seshadri Nadathur}
\affiliation{Institute of Cosmology and Gravitation, University of Portsmouth, Dennis Sciama Building, Portsmouth, PO1 3FX, UK}
\email{seshadri.nadathur@port.ac.uk}

\author[0000-0002-0644-5727]{Will Percival}
\affiliation{Department of Physics and Astronomy, University of Waterloo, 200 University Ave W, Waterloo, ON N2L 3G1, Canada}
\affiliation{Perimeter Institute for Theoretical Physics, 31 Caroline St. North, Waterloo, ON N2L 2Y5, Canada}
\affiliation{Waterloo Centre for Astrophysics, University of Waterloo, 200 University Ave W, Waterloo, ON N2L 3G1, Canada}
\email{will.percival@uwaterloo.ca}

\author{Claire Poppett}
\affiliation{Lawrence Berkeley National Laboratory, 1 Cyclotron Road, Berkeley, CA 94720, USA}
\affiliation{Space Sciences Laboratory, University of California, Berkeley, 7 Gauss Way, Berkeley, CA  94720, USA}
\affiliation{University of California, Berkeley, 110 Sproul Hall \#5800 Berkeley, CA 94720, USA}
\email{clpoppett@lbl.gov}

\author[0000-0001-6979-0125]{Ignasi P\'erez-R\`afols}
\affiliation{Departament de F\'isica, EEBE, Universitat Polit\`ecnica de Catalunya, c/Eduard Maristany 10, 08930 Barcelona, Spain}
\email{ignasi.perez.rafols@upc.edu}

\author[0000-0001-7145-8674]{Francisco Prada}
\affiliation{Instituto de F\'{i}sica Te\'{o}rica (IFT) UAM/CSIC, Universidad Aut\'{o}noma de Madrid, Cantoblanco, E-28049, Madrid, Spain}
\affiliation{Instituto de Astrof\'{i}sica de Andaluc\'{i}a (CSIC), Glorieta de la Astronom\'{i}a, s/n, E-18008 Granada, Spain}
\affiliation{Instituto de Astrof\'{\i}sica de Canarias, C/ V\'{\i}a L\'{a}ctea, s/n, E-38205 La Laguna, Tenerife, Spain}
\email{fprada@iaa.es}

\author{Graziano Rossi}
\affiliation{Department of Physics and Astronomy, Sejong University, 209 Neungdong-ro, Gwangjin-gu, Seoul 05006, Republic of Korea}
\email{graziano@sejong.ac.kr}

\author[0000-0002-9646-8198]{Eusebio Sanchez}
\affiliation{CIEMAT, Avenida Complutense 40, E-28040 Madrid, Spain}
\email{eusebio.sanchez@ciemat.es}

\author[0000-0002-3569-7421]{Edward Schlafly}
\affiliation{Space Telescope Science Institute, 3700 San Martin Drive, Baltimore, MD 21218, USA}
\email{eschlafly@stsci.edu}

\author{David Schlegel}
\affiliation{Lawrence Berkeley National Laboratory, 1 Cyclotron Road, Berkeley, CA 94720, USA}
\email{djschlegel@lbl.gov}

\author{Michael Schubnell}
\affiliation{Department of Physics, University of Michigan, 450 Church Street, Ann Arbor, MI 48109, USA}
\affiliation{University of Michigan, 500 S. State Street, Ann Arbor, MI 48109, USA}
\email{schubnel@umich.edu}

\author{David Sprayberry}
\affiliation{NSF NOIRLab, 950 N. Cherry Ave., Tucson, AZ 85719, USA}
\email{david.sprayberry@noirlab.edu}

\author[0000-0003-1704-0781]{Gregory Tarl\'e}
\affiliation{University of Michigan, 500 S. State Street, Ann Arbor, MI 48109, USA}
\email{gtarle@umich.edu}

\author{Benjamin Alan Weaver}
\affiliation{NSF NOIRLab, 950 N. Cherry Ave., Tucson, AZ 85719, USA}
\email{benjamin.weaver@noirlab.edu}

\author[0000-0001-5381-4372]{Rongpu Zhou}
\affiliation{Lawrence Berkeley National Laboratory, 1 Cyclotron Road, Berkeley, CA 94720, USA}
\email{rongpuzhou@lbl.gov}

\author[0000-0002-6684-3997]{Hu Zou}
\affiliation{National Astronomical Observatories, Chinese Academy of Sciences, A20 Datun Road, Chaoyang District, Beijing, 100101, P. R. China}
\email{zouhu@nao.cas.cn}

\collaboration{all}{The Dark Energy Spectroscopic Instrument collaboration}


\begin{abstract}

We present the first systematic spectroscopic observations of extragalactic transients from the Dark Energy Spectroscopic Instrument (DESI), as part of the DESI Transients Survey program. With 5,000 fibers and an ${\sim} 8$ deg$^2$ field of view, we exploit DESI as a machine for the discovery and classification of transients. We present transient classifications from archival DESI data in Data Releases 1 and 2, relying on a combination of a secondary target program and serendipitous observations. We also present observations from the first 6 months of the DESI spare fiber program dedicated to transients. The program is run in coordination with a dedicated DECam time-domain survey, serving as a pathfinder for what we will be able to achieve in conjunction with the Rubin Observatory Legacy Survey of Space and Time (LSST).
We classify over 250 transients, of which the majority were previously unclassified. The sample comprises thermonuclear and core-collapse supernovae and tidal disruption events (TDEs), including a TDE observed before its discovery in imaging. We demonstrate DESI's ability to classify a population of faint transients down to $r\sim 22.5$ mag during main survey operations, with negligible impacts on DESI's main observations. With the start of Rubin LSST operations, we expect to classify $\mathcal{O}(1000)$ transients per year.

\end{abstract}

\keywords{\uat{Supernovae}{1668} --- \uat{Core-collapse Supernovae}{304} --- \uat{Type Ia Supernovae}{1728} --- \uat{Transient detection}{1957} --- \uat{Transient sources}{1851} }


\section{Introduction}

During the past decade, wide-field imaging surveys such as the Zwicky Transient Facility (ZTF; \citealt{bellm_zwicky_2018, masci_zwicky_2018, bellm_zwicky_2019, graham_zwicky_2019, dekany_zwicky_2020}), the Asteroid Terrestial Last-Alert System (ATLAS; \citealt{tonry_early_2010, tonry_atlas_2018, smith_design_2020}), the All-Sky Automated Survey for Supernoave (ASAS-SN; \citealt{shappee_all_2014}), and the Panchromatic Survey Telescope and Rapid Response System (PanSTARRS; \citealt{kaiser_pan-starrs_2002}), have proven instrumental in the discovery of tens of thousands of extragalactic transients. These surveys stream public alerts or periodically upload interesting transients to databases such as the Transient Name Server (TNS; \citealt{gal-yam_tns_2021}), allowing the entire astronomical community to become aware of new discoveries promptly. However, the large number of transients reported to TNS is of little scientific value without robust classifications, generally provided by spectroscopy. Currently, only ${\sim}12\%$ of the transients reported to TNS have a classification. Although significant advances have been made in photometric classification (e.g. \citealt{campbell_cosmology_2013, moller_photometric_2016, vincenzi_dark_2024}), a large number of transients reported to TNS do not have publicly available photometry (${\sim}40\%$). Furthermore, photometric classification cannot reach the level of certainty provided by spectroscopy.

Some programs have made immense efforts to spectroscopically classify a high fraction of the numerous discovered transients, such as the Bright Transient Survey (BTS; \citealt{fremling_zwicky_2020, perley_zwicky_2020, rehemtulla_zwicky_2024}), the Census of the Local Universe (CLU; \citealt{cook_census_2019, de_zwicky_2020}), the Public ESO Spectroscopic Survey of Transient Objects (PESSTO; \citealt{smartt_pessto_2015}), and the Complete Astronomical Transient Survey within 150 Mpc (CATS-150; \citealt{das_ztf_2025}). However, these surveys typically use hard cuts in apparent magnitude or distance (BTS; $m < 18.5$ mag, CLU; $d_L < 50$ Mpc, PESSTO; $m < 20.5$ mag, CATS-150; $d_L < 150$ Mpc). Thus, a large number of fainter transients at higher redshifts remain unclassified, limiting our ability to trace their evolution with redshift and probe their progenitors over cosmic time. Pushing the redshift limits of spectroscopic classification allows us to look back at a younger universe. It connects directly to progenitors because the defining features of a supernova (SN) spectrum are directly related to the composition and stripping state, providing information on mass loss and binary evolution \citep{smartt_progenitors_2009}.

The importance of spectroscopic classification cannot be understated. For example, detailed spectroscopic studies of core-collapse supernovae (CCSNe) have led to a better understanding of stellar evolution and binary star mass loss \citep{schmidt_unusual_1993}. Early discovery with wide-field surveys, allows for both infant and late time spectra of CCSNe. Such spectra provide information on the nebular interaction of a SN with their surrounding environment, allowing to probe progenitor mass loss and environment \citep{fransson_late_1989, jerkstrand_progenitor_2012, hueichapan_optical_2025}. CCSNe have also led to fascinating multi-messenger results, such as the detection of neutrinos from SN 1987A \citep{hirata_observation_1987}. Furthermore, CCSNe have been robustly associated to long duration gamma-ray bursts through spectroscopic confirmation \citep[e.g.,][]{galama_unusual_1998}. More recently, CCSNe have also begun to be connected to a different class of extragalactic fast X-ray transients \citep[EFXTs;][]{sakamoto_global_2005, heise_x-ray_2001, negoro_maxigsc_2016, srinivasaragavan_ep250827bsn_2025}.

Thermonuclear SNe are an instrumental standard candle to measure cosmological distances \citep{goobar_supernova_2011}, which led to the incredible discovery of the accelerating expansion of the universe \citep{riess_observational_1998, perlmutter_measurements_1999}. The cosmological distances probed by SNe also led to one of the most prominent stress tests of the $\Lambda$CDM model through the Hubble Tension \citep[see review by][]{valentino_cosmoverse_2025}. While much effort has gone into the pure photometric identification and modeling of Type Ia SNe \citep{moller_photometric_2016, moller_dark_2022, mitra_fully_2025}, spectroscopy can still improve the standardization process \citep{bailey_using_2009, blondin_spectra_2011,Boone_2021}. Type Ia SNe that are potentially produced by white dwarf mergers will also offer a new type of standard candle with the Laser Interferometer Space Antenna \citep{nelemans_gravitational_2001}. 

Tidal disruption events (TDEs) are a class of nuclear transients that occur when a star passes too close to a massive black hole (MBH), which shreds the star and rapidly accretes the disrupted gas. \citep{hills_possible_1975, gezari_tidal_2013}. TDEs offer a unique opportunity to measure the masses and spins of dormant MBHs, and to study stellar populations and dynamics in galactic nuclei \citep{velzen_seventeen_2021, hammerstein_final_2022, yao_tidal_2023, mockler_uncovering_2023, graur_dependence_2018}. These events are exceedingly rare, with ${\sim}240$ total discovered to date \citep{franz_open_2025}.  

With the recent start of the Vera C. Rubin Observatory Legacy Survey of Space and Time \citep[LSST;][]{ivezic_lsst_2019}, the difficult task of comprehensive spectroscopic investigation is about to become even more challenging. Rubin is expected to discover on the order of 100,000 transients per night, nearly 100 times the rate of ZTF. Faced with an enormous number of potential targets, many groups have developed algorithms and machine learning models to help identify the most interesting transients that will be discovered by Rubin (e.g. \citealt{boone_parsnip_2021}). Several of these classifiers exist with an assortment of different goals. 

The first type are alert based classifiers, these use metadata and cutouts from single alerts to identify targets of interest. For example, \texttt{BTSBot} aims to identify all SNe that will reach a specific brightness criterion \citep{rehemtulla_zwicky_2024}. The Automatic Learning for the Rapid Classification of Events (ALeRCE) stamp classifier attempts to offer a quick prediction of the  transient type \citep{forster_automatic_2021, carrasco-davis_alert_2021}. Finally, there exist classifiers that use both alert and light-curve data to attempt to recover targets of unique interest such as TDEs and superluminous SNe; these include \texttt{TDEScore} \citep{stein_tdescore_2024}, \texttt{NEEDLE} \citep{sheng_neural_2024}, ALeRCE's Lightcurve Classifier \citep{sanchez-saez_alert_2021, pavez-herrera_alerce_2025}, and \texttt{FLEET} \citep{gomez_first_2023, gomez_identifying_2023}. However, these programs will still inevitably recommend more suitable targets than there is spectroscopic follow-up capability in the era of Rubin LSST.

To aid in expanding the follow-up capability to discover these fascinating transients, we have begun investigating the use of the Dark Energy Spectroscopic Instrument (DESI) as a transient discovery and classification machine. DESI is a robotic fiber-fed multi-object spectrograph mounted on the Mayall 4m telescope at Kitt Peak National Observatory (KPNO) and is designed to obtain spectra of over 60 million galaxies over eight years of operation \citep{desi_collaboration_desi_2016-1, desi_collaboration_desi_2016}. In a single night, DESI is able to observe 100,000 different galaxies, and can therefore, on its own, discover a large number of transients. 

In this work, we demonstrate the ability to use DESI as a transient discovery and classification machine, supplementing the use of DESI in the characterization of supernova host galaxies \citep{2024ApJS..275...22S}, as well as of the environments and large scale structure around gravitational wave events \citep{2019BAAS...51c.310P,Palmese:2021mjm,2023RNAAS...7..250B,amsellem2025probing,2025arXiv251023723H}. In Section \ref{sec:data} we describe DESI and the Dark Energy Camera (DECam) Survey for Intermediate Redshift Transients (DESIRT). In Section \ref{methodology}, we outline how transients were identified in DESI and how targets for the ongoing DESI Transients Survey, a spare fiber program, are selected from publicly reported transients and DESIRT. In Section \ref{classification}, we explain the methodology for classifications that will be reported on the Transient Name Server (TNS). Finally in Section \ref{results}, we present the first classifications of 163 transients, and confirm the classifications of 58 transients observed by DESI based on the data obtained in Data Release 2 (DR2; May 2021 - June 2024; \citealt{DESI2024.VII.KP7B, desi_collaboration_data_2025, desi_collaboration_desi_2025}). This sample represents a sizable data set of high-quality SN spectra, including those of several infant SNe. We show how TDEs can be studied with these observations. Finally, we demonstrate that our survey targets a less investigated population of transients at higher redshifts and fainter magnitudes than systematically done with previous spectroscopic surveys, and present our conclusions in Section \ref{sec:conclusions}.

Throughout this work, we assume a flat $\Lambda$CDM cosmology with $H_0=67.66~$ km s$^{-1}$ Mpc$^{-1}$ and $\Omega_m=0.31$ \citep{aghanim_planck_2020}. All magnitudes are reported in the AB system.

\section{Data}\label{sec:data}

\subsection{DESI}
\label{sec:desi}
DESI is comprised of 5,000 fibers that can each be positioned at a unique place on sky \citep{poppett_overview_2024}. The DESI instrument covers an ${\sim}8$ square degree area on the sky \citep{Corrector.Miller.2023}. Each fiber is 107 $\mu$m in diameter, which is equivalent to an angular on-sky diameter of ${\sim}1.5^{\prime\prime}$. The spectrograph is made up of three cameras (BRZ) that cover $3600-9824~\text{\AA}$ at a R $\sim 2000 - 5500$ \citep{desi_collaboration_desi_2016-1, desi_collaboration_desi_2016, DESI2022.KP1.Instr}\footnote{DESI Collaboration Key Paper}. Each spectrum is reduced and flux calibrated with the DESI spectroscopic data pipeline \citep{guy_spectroscopic_2023}. To establish a redshift from a given DESI spectrum, template spectra are fit to the target using \texttt{redrock}\footnote{\url{https://github.com/desihub/redrock/}} and a redshift is determined via $\chi^2$ minimization \citep[see][]{Redrock.Bailey.2024}. Unfortunately, this can lead to catastrophic failure when dealing with transient spectra if there is insufficient host contamination (i.e., a lack of host galaxy emission and absorption features) for the galaxy-based template to fit a redshift. 

The DESI Survey is comprised of three main tiling programs `bright', `dark', and `backup'. The backup program mainly focuses on stars and thus is not relevant to our analysis. The `bright' and `dark' programs are defined by their target classes and sky conditions, with dark exposures ranging from $900-1200$s integration time and bright exposures ranging from $150-300$s total exposure \citep[see][]{schlafly_survey_2023}.

\subsection{The DECam Survey for Intermediate Redshift Transients}
\label{sec:desirt}

The DECam Survey for Intermediate Redshift Transients (DESIRT; PIs: Palmese \& Wang) is a wide field survey that observes ${\sim}100$ square degrees of sky in a combination of the $griz$ bands down to a $5\sigma$ limit of $r \approx 23.5$ mag with a cadence of $3$ days \citep{palmese_desirt_2022}. DESIRT has the goal of producing high-quality light curves for thousands of extragalactic transients at a $z > 0.2$ or a peak brightness of $r\approx20.5$ mag. This survey targets equatorial fields that are expected to be observed by DESI within the same weeks or months. Note that DESI's observations are scheduled on the-fly throughout the night depending on the observing conditions, thus it is not possible to predict exactly where DESI will observe, although we have an idea of which tiles are more likely to be observed around a given time. This synergy also allows for any serendipitous DESI transients to have well-defined and observed lightcurves. The first stage of DESIRT ran from 2021-2023 and a similar program was started in February 2025 as the DECam DESI Transients Survey (PI: Palmese; \citealt{hall_decam_2025}). 

DESIRT uses the Saccadic Fast Fourier Transform (SFFT) algorithm developed by \citet{hu_image_2022} to enable fast and accurate difference imaging to identify transient candidates and produce alerts, see \citet{Cabrera_2024} and \citet{Hu2025} for further details on the full analysis pipeline. The transient alerts are then processed with a real-bogus convolution neural network to reject artifacts such as cosmic rays (Hu et al., in prep.). The pipeline then performs a cross-match of the alerts to Gaia DR3 to remove known stars \citep{vallenari_gaia_2023}. Finally, a match to the Legacy Survey star-galaxy catalog \citep{liu_morphological_2025} is performed to remove any remaining stellar alerts based on an archival source’s morphology in Legacy Survey imaging \citep{dey_overview_2019}. The remaining transients are then packaged into Alerts and sent out in a Kafka stream. The hand-selected transients, based on a visual lightcurve inspection by X.J.H., from this survey are then reported to TNS.

\section{Discovery and Follow-up Methodology} \label{methodology}

\subsection{Serendipitous Observations}

As DESI observes over $100,000$ different galaxies per night, it has inevitably serendipitously observed numerous nuclear transients. By both spatially and temporally matching all DESI observations with all probable supernovae (PSN) from TNS, we can quickly determine all serendipitous transient observations by DESI. We then seek to classify these transients as outlined in subsequent sections of this manuscript.

In order to select these events, we match all transients that occurred within a $0.75^{\prime\prime}$ radius of a DESI observation within DR1 or DR2 \citep{, desi_collaboration_data_2025, desi_collaboration_desi_2025} \footnote{DESI Collaboration Key Paper}. This number was chosen due to the full diameter of a DESI fiber being $1.5^{\prime\prime}$ (see \S \ref{sec:desi}). Thus, in order to capture these transients, the source must positionally occur inside the DESI fiber. We note, however, that because of the telescope’s PSF, light from a transient located outside the physical fiber radius can still bleed into the aperture \citep{graur_discovery_2013}. Finding such instances would require a larger search radius and increase the number of candidates. This proximity cut means that most serendipitous transients are located near the nucleus of their likely host galaxy. In theory, there could be exceptions to this through various secondary programs (see, e.g., Section 2.1 of \citealt{myers_target-selection_2023}) or failures in the Legacy Survey tractor catalog incorrectly identifying one large (potentially clumpy) galaxy as many small galaxies \citep{dey_overview_2019}. However, in our sample, all serendipitous targets were located in a fiber placed in a nuclear position.

We search for any reported transients on TNS that were observed by DESI either 30 days before the discovery date or up to 120 days after the discovery date. This covers most of the time range over which we may expect to detect a standard SN at redshift $\lesssim 0.5$ \citep{filippenko_optical_1997}. This also provides plenty of time for slower events like TDEs \citep{bricman_rubin_2023}. 

\begin{figure}[h]
    \centering
    \includegraphics[width=1.02\linewidth]{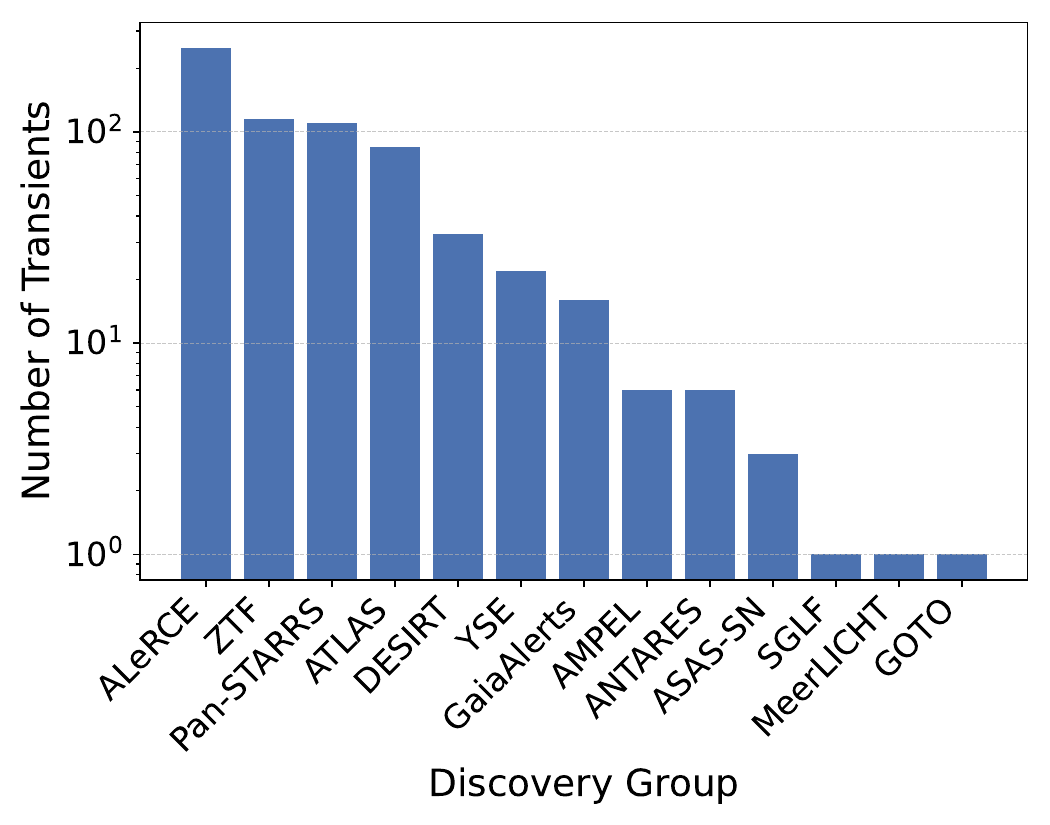}
    \caption{TNS transients that have serendipitous observations by DESI per reporting group. Transients come from ZTF (ALERCE \citealt{sanchez-saez_alert_2021}, ZTF Partnerships \citealt{bellm_zwicky_2018}, and AMPEL \citep{nordin_transient_2019}), Pan-STARRS \citep{chambers_pan-starrs1_2016}, ATLAS \citep{tonry_atlas_2018}, DESIRT \citep{palmese_desirt_2022, hall_decam_2025}, The Young Supernova Experiment \citep[YSE;][]{jones_young_2021}, }
    \label{fig:discovery_group}
\end{figure}

Using these criteria, we selected 649 transients with observations contained in DR2 \citep{desi_collaboration_data_2025} that are determined to have DESI spectra observed within the time period of potential transient activity, 30 days before until 120 days after discovery. In Figure \ref{fig:discovery_group} we show the breakdown of surveys reporting these 649 transients.

\subsection{Spare Fiber Program}

\begin{figure*}
    \centering
    \includegraphics[width=\textwidth]{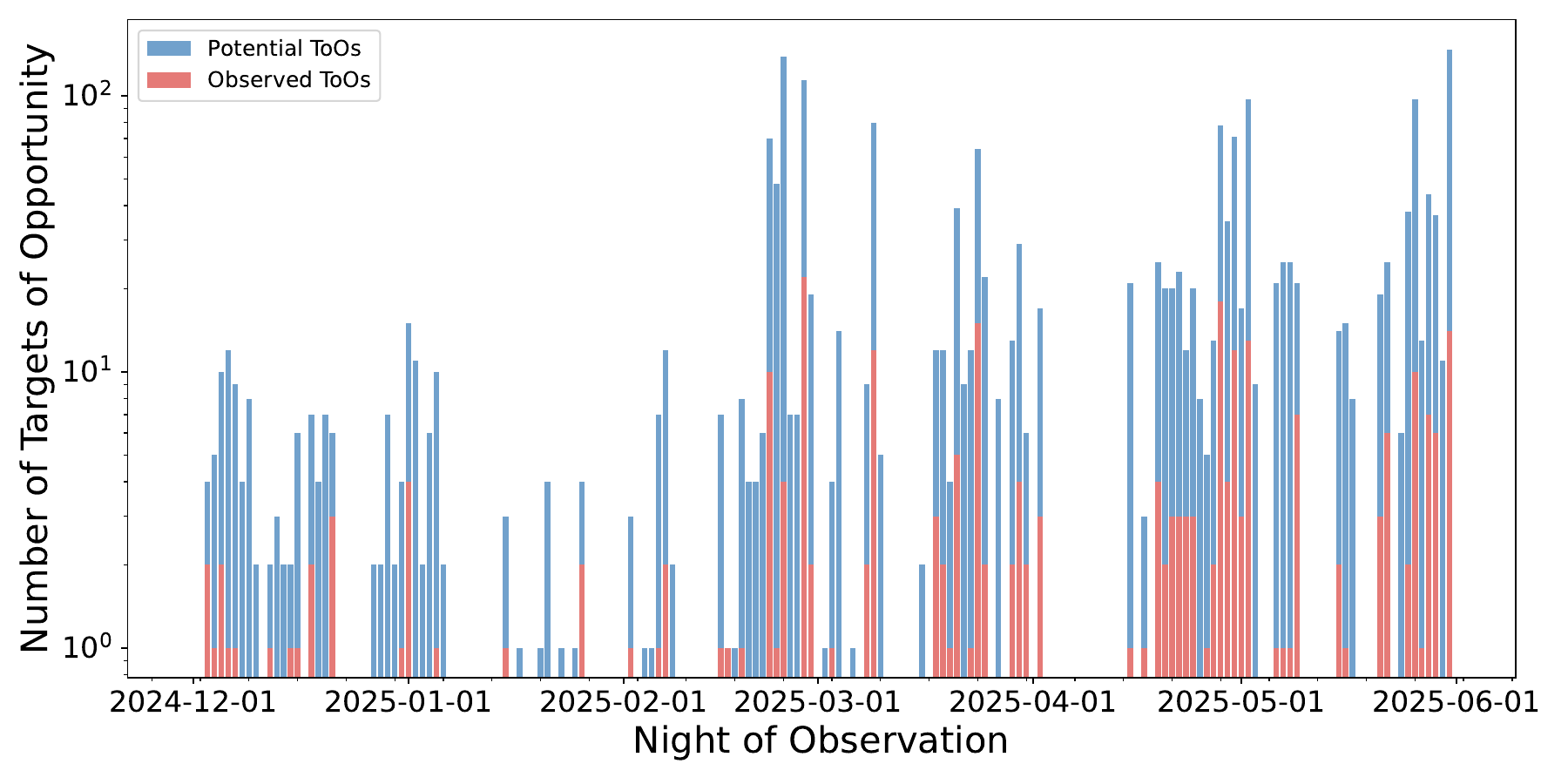}
    \caption{The number of targets of opportunity that fall into a DESI tile on a given survey night (Potential ToOs) versus the number actually observed (Observed ToOs). On nights where DARK or BRIGHT tiles are observed about $10\%$ of all possible targets are observed. We find that DESI is able to average about one transient per night from TNS alone and around three per night including TNS, ZTF, and DESIRT. This large jump in observed and potential ToOs can be seen with the restarting of the DESIRT program in February 2025.}
    \label{fig:too_counts}
\end{figure*}

To improve the classification completeness of public transient datasets available on TNS, as well as to drive the discovery of new kinds of transients identified by public wide field surveys, we have implemented a spare-fiber program with DESI to provide spectroscopic classifications for transients that may have been unprioritized by other programs \citep[e.g.,][]{DESI:2023ytc}. DESI has 5,000 different fibers that can each be assigned to a different target in the sky. Over different exposures at any given pointing these fibers move to observe a different set of targets (a ``tile''). However, due to the large scale structure distribution of galaxies and the need to avoid fiber collisions, some fibers have no target available to be assigned. Although some fibers are needed for sky observations to be used in the calibration steps, typically only $\sim3,000-4,000$ DESI main targets are pointed to in each observation. This leaves hundreds of fibers available per tile. During Survey Validation (SV) and the first years of the DESI main survey, secondary target programs have been run to make use of these spare fibers \citep[e.g.,][]{DESI:2023ytc}. See \citet{Myers:2022azg,DESI:2023mkx} for a larger description of the secondary targeting strategy. Thus, by adding transient targets of opportunity (ToO), we can utilize these otherwise spare fibers to obtain useful spectroscopic observations and classifications.


A first version of the DESI Transients Survey was run as a secondary target program during the SV phase and part of the main survey, mostly only feeding in DESIRT transients discovered using DECam (see \S \ref{sec:desirt} and Appendix B of \citealt{desi_collaboration_early_2024}). In late November 2024, we started the DESI Transients Survey, a spare fiber program that targets publicly reported transients following a set of criteria. This new version of the DESI transients Survey will serve as a pathfinder for future observations of Rubin transients with DESI.

The current method for determining targets for the DESI Transients Survey proceeds as follows:
\begin{enumerate}
  \item All transients classified as probable supernova, labeled as ``PSN", on TNS with at least one observation in the past week brighter than 22nd magnitude in any filter. 
  \item All ZTF alerts in the past week that pass the selection cuts outlined in \citet{perley_zwicky_2020} and \citet{rehemtulla_zwicky_2024}. 
  \item All transients discovered by DESIRT with observations or hosts brighter than 22nd magnitude in the last week in any filter. 
\end{enumerate}

These targets are then submitted to a candidate list and positioned by the DESI on-the-fly fiberassign software.\footnote{\url{https://github.com/desihub/fiberassign}}

Currently, the fiber assign code takes two priorities, \texttt{HI}, which refers to a priority higher than main survey targets, and \texttt{LO}, which has a priority lower than main survey targets and only higher than ``filler" programs \citep{myers_target-selection_2023}. The higher priority is currently disabled for main survey tiles and only used for special tiles such as calibration tiles. ToO targets can be assigned to either the bright or dark program if they are assigned to the \texttt{BRIGHT} program, or only by the dark program if they are entered as \texttt{DARK} targets. Each bright tile has a $180$ s effective exposure time and is generally taken during bright moon phases and towards dawn or dusk. Dark tiles have a $1000$ s effective exposure time and are taken under ideal conditions with less than 60\% moon illumination \citep{DESI:2023mkx}.


\begin{figure}[h]
    \centering
    \includegraphics[width=1.02\linewidth]{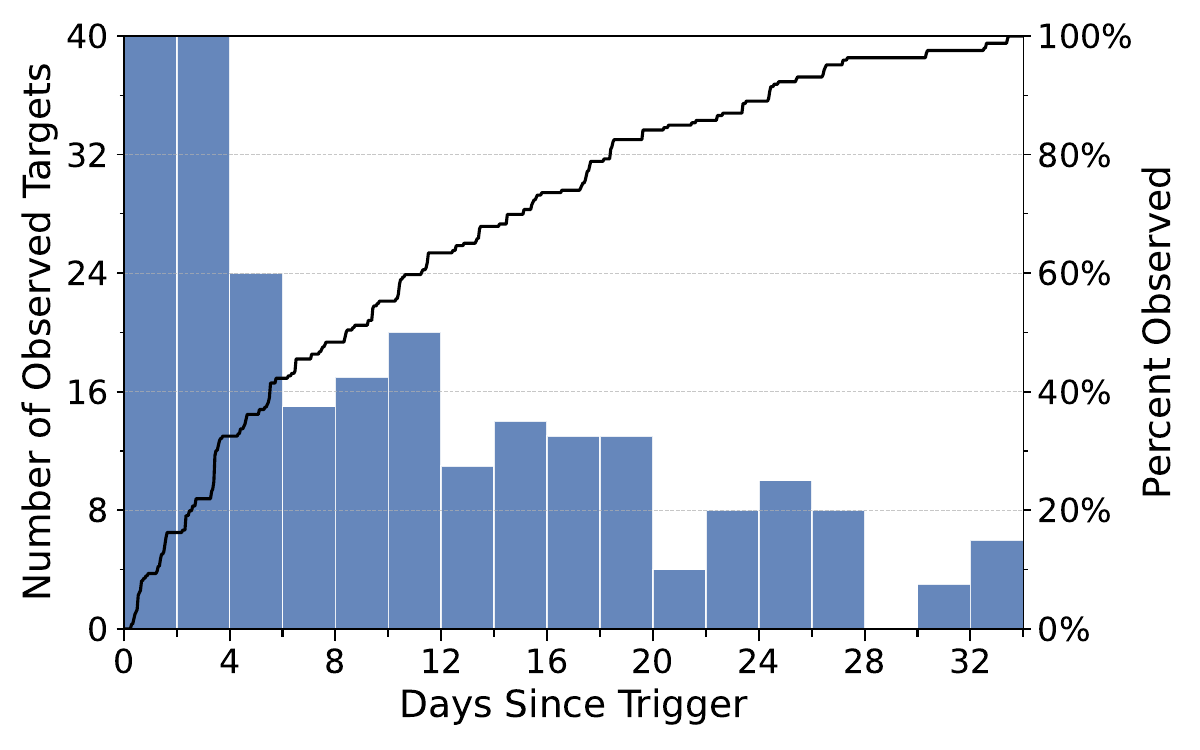}
    \caption{A histogram of the number of transients that have been observed a certain number of days after being initially triggered. The line shows the cumulative distribution of targets observed within that time span.
    }
    \label{fig:timeframe}
\end{figure}

As can be seen in Figure \ref{fig:too_counts}, the number of ToOs observable by DESI by combining these different targeting sources is on the order of $10 - 50$ each night. On average during the first six months of the spare fiber program (1 December 2024 - 1 June 2025), DESI has been able to observe $13\%$ of the TNS transients that were submitted to the program and that ended up in an unobserved DESI tile. In these first six months, we have also demonstrated DESI's ability to quickly follow-up transients. We find that ${\sim}50\%$ of transients observed, are observed within $8$ days of being submitted as a target of opportunity (Figure \ref{fig:timeframe}).



\section{Classification Process} \label{classification}

\subsection{Classifying DESI Spectra}

The classification of transients discovered by DESI poses unique challenges unlike those of more standard follow-up campaigns. First, the serendipitous nature of the discovery of a transient means that there exists generally minimal photometric information regarding the targets. Many of these transients peak near the limiting magnitudes of the discovery survey, which is why no pre-existing classification exists for many of these targets. Their faint nature is largely due to the higher redshift distribution of these transients.

Although targets submitted through the ToO spare-fiber program can be at any location with respect to their host galaxy, many of the serendipitously observed DESI transient spectra are from fibers placed in the nuclear region of the galaxies, which results in significant contamination from the host. That, combined with the high resolution and sensitivity of the spectrograph, means that narrow galactic emission lines not usually found in SN spectra are almost always present. Thus, normal spectral extragalactic classifiers such as SNID and DASH \citep{blondin_determining_2007, muthukrishna_dash_2019} struggle to classify these targets commonly believing that the narrow lines are indicative of a Type IIn SNe which demonstrate narrow Hydrogen emission lines or Ia-CSM which also displays very narrow hydrogen emission lines \citep[e.g.,][]{fox_nature_2015, ransome_systematic_2021, sharma_systematic_2023}. 

\subsubsection{Next Generation Superfit}

For classification, we use a modified, multithreaded version of Next Generation Superfit \citep[NGSF;][]{goldwasser_next_2022, howell_gemini_2005}\footnote{\url{https://github.com/Hallflower20/superfit}}. NGSF works particularly well for this situation due to its simultaneous fitting of both galaxy and transient templates, a methodology highlighted in \citet{graur_discovery_2013, graur_unified_2015}. It also estimates a percentage of the spectral flux produced by the transient. If this percentage is significantly low ($< 20\%$), it is possible that the transient classification is not reliable. 

The DESI pipeline \texttt{redrock} fits galaxy templates to galaxy spectra to measure redshifts \citep{Redrock.Bailey.2024}. However, the broad lines produced by supernovae can often be misinterpreted in these template fits. Thus, NGSF was run with two different settings. An initial run with a fixed redshift and the galaxy lines masked, with the intent that if there are strong narrow galaxy lines, the redshift will still be correctly assigned by \texttt{redrock}. The second setting turns off the fixed redshift and allows NGSF to pick the best redshift from the best matching template. This works better for situations where the DESI redshift may be incorrect, severely off due to misidentification of supernova features, or in the case of off nuclear fiber placement that leads to a significant lack of host light.

\subsubsection{Human Vetting}

Upon fitting the NGSF template, each template and spectrum was plotted and inspected. Furthermore, important broad SN lines (H, He, Si, S, Ca, Fe, etc.) were investigated using the DESI Spectra Viewer\footnote{\url{https://www.legacysurvey.org/viewer/desi-spectrum/dr1/targetid39633493369555948}}. The best matching template's redshift was identified, and the template's broad classification was assigned using both the manual identification of broad lines and the NGSF recommendation. However, in many cases there simply appeared to be no clear transient signature present and the DESI spectrum inspected only showed host galaxy emission. If the transient contribution was estimated to be less than $20\%$ by NGSF then no transient classification was made. 

\subsection{Class Breakdown}

We follow a classification structure similar to that described in \citet{fremling_zwicky_2020} which makes use of the broad classes SNe Ia, SNe II, and SNe Ib/c. Furthermore, we classified any potential tidal disruption events discovered. These classifications are made almost entirely through a spectral analysis. Thus, we did not assign a Super-Luminous classification to any transient. The lack of high-cadence photometric observations for these relatively dim SNe makes it difficult to determine a peak brightness to meet the criteria (absolute magnitude M $< -21$ mag) as outlined by \citet{gal-yam_luminous_2012}.

\section{Results} \label{results}

\subsection{Legacy Classifications}

Of the 853 legacy spectra inspected, 649 of which are serendipitous BGS targets and 204 are secondary targets or legacy ToO targets, only 198 had a clearly identifiable SN or TDE spectrum. The unclassifiable spectra are dominated by host emission or were taken too early or late to effectively capture a transient spectrum. Both of these issues make it difficult to give a confident identification. 48 of these already have classifications on TNS. We use these to confirm our method; we find that there are 29 SNe Ia, 14 SNe II, 2 SNe Ib/c, and 3 TDEs. We find that of the 150 newly classifiable transients, 148 are SNe and 2 are TDEs. Of the 148 SNe, 123 are SNe Ia, 22 are SNe II, and 3 are SN Ib/c. 

From December 1st 2024 to May 1st 2025, 184 spectra were taken with the spare fiber ToO program. Of these, 59 are classified as 44 SNe Ia, 11 SNe II, and 4 SNe Ib/c. The complete classification breakdown can be seen in Figure \ref{fig:type_treemap}. Moreover, we show examples of our classied spectra in Appendix \ref{app}, which also includes a time series observation of SN 2023ixf (Appendix \ref{app}, Figure \ref{fig:SN2023ixf}; \citealt{2023TNSAN.137....1B}), one of the closest Type II SN ever observed (e.g. \citealt{Zimmerman_2024}). Our breakdown of novel SNe classifications closely matches the ratios reported in \citet{fremling_zwicky_2020}, but, stripped envelope SNe seem underrepresented. A Poisson estimate of uncertainties gives $7.3\%\pm1.0\%$ versus the observed $3.5\%\pm1.3\%$, thus the difference is about $2.4\sigma$ which is insufficient to claim a rate discrepancy, but may point to some systemic differences in the survey methods becoming apparent. 

\begin{figure}
    \centering
    \includegraphics[width=1\linewidth]{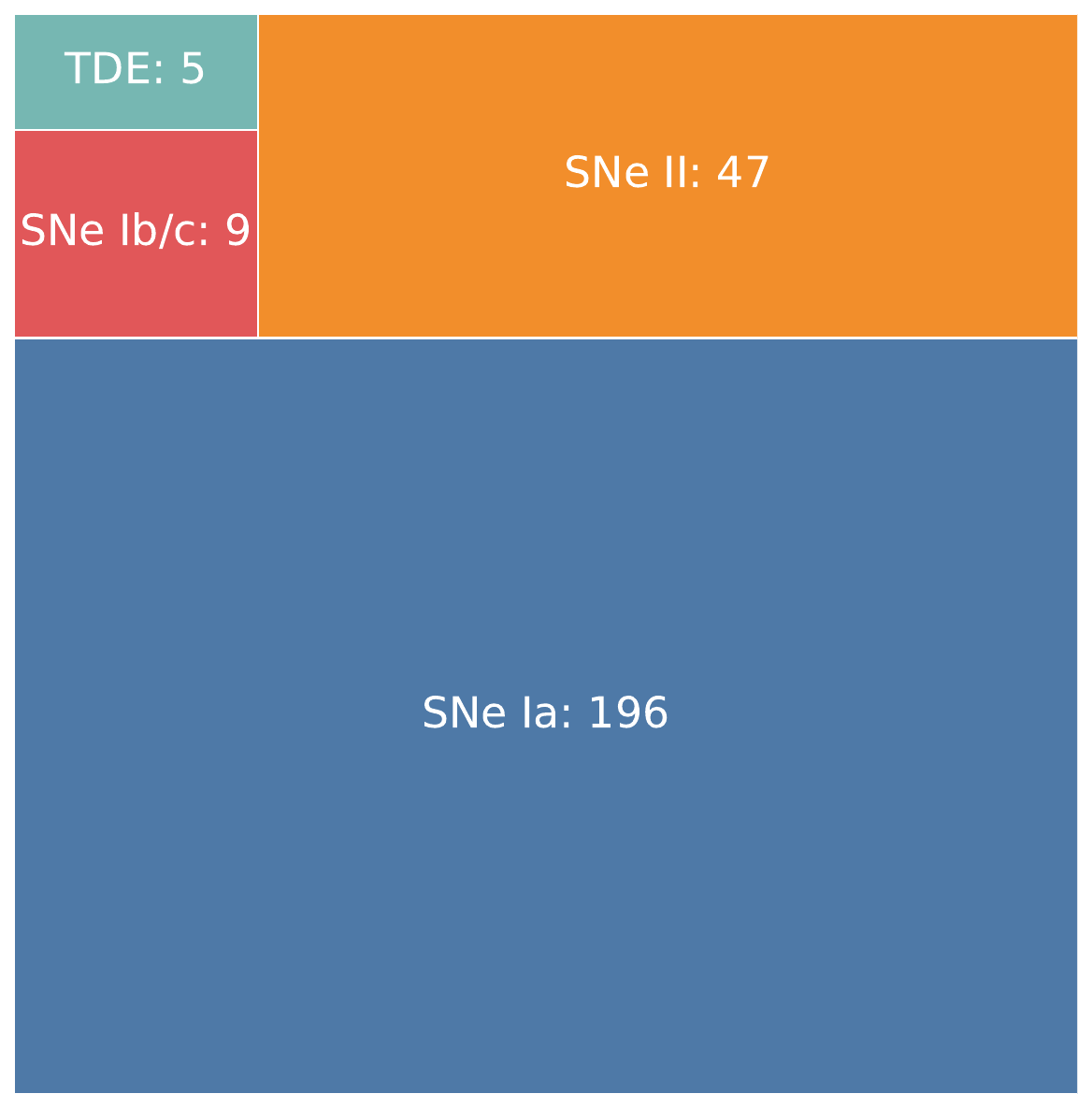}
    \caption{Treemap showing the count of each broad type of extragalactic transients discovered in DESI DR1, DR2 and the spare fiber survey \citep{desi_collaboration_data_2025, desi_collaboration_desi_2025}}
    \label{fig:type_treemap}
\end{figure}

\subsection{Tidal Disruption Events}

TDEs are characterized by light curves with long rise times of ${\sim}30$ days. They also generally maintain a constant blue color with little reddening over time. Their spectra resemble black bodies with broad hydrogen or helium features. However, in novel cases of especially bright TDEs there is a new featureless class \citep{hammerstein_final_2022}. In the following, we present each of the newly classified TDEs in DESI DR1 and DR2  \citep{desi_collaboration_data_2025, desi_collaboration_desi_2025}.

\begin{figure*}[h]
    \centering
    \includegraphics[width=0.98\textwidth]{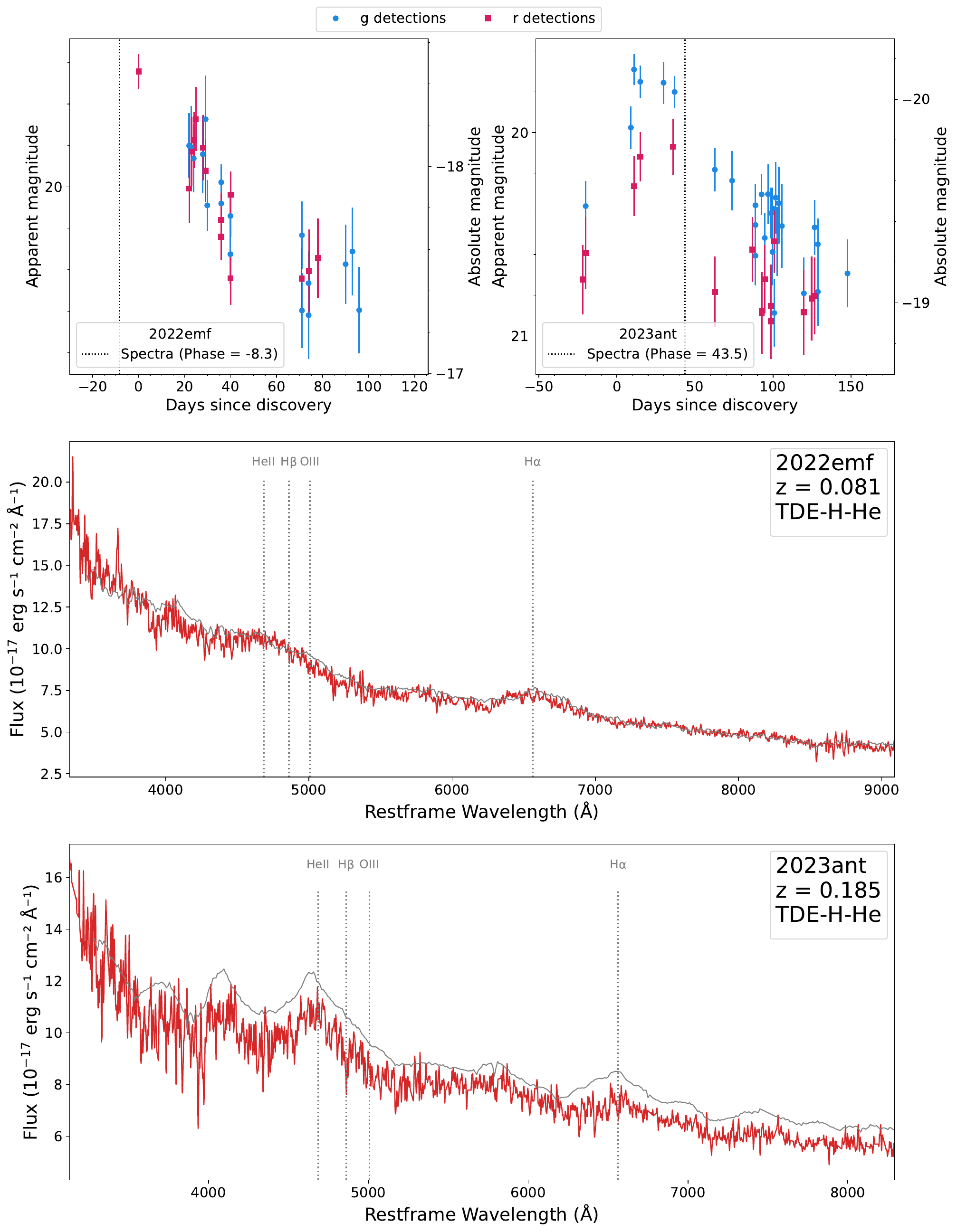}
    \caption{\emph{Top:} lightcurves of TDEs 2022emf (left) and 2023ant (right), two TDEs newly classified by this work; in both note the long slow decline and lack of clear color evolution in the lightcurve. \emph{Bottom:} DESI spectra of the same TDEs with the ``best fit" template from Superfit is plotted in gray. The broad He II and H$\alpha$ emission in each spectrum are consistent with TDEs. The vertical line shows the time of the DESI spectrum. The spectrum for 2022emf was taken even before the transient discovery in imaging. DESI serendipitous and spare fiber spectroscopy has immense potential in its ability to do infant TDE and SNe studies. The lightcurves are provided by the ZTF forced-photometry service \citep{masci_new_2023}.}
    \label{fig:tde_data}
\end{figure*}

\subsubsection{TDE 2022emf}

2022emf was first reported by a ZTF alert as a transient on 2022-01-21 and was observed by DESI on 2022-01-14. Thus, our spectrum was taken seven days before any public report. We classify 2022emf as a TDE-H-He. We give this classification due to the broad H$\alpha$ emission present in the spectrum as well as the blackbody of the spectrum. We provide an early time spectrum observed 9 days before the first ZTF observation which is also the peak in the ZTF observations. The light curve shows a slow decay (Figure \ref{fig:tde_data}); unfortunately, the color evolution is difficult to determine because the observations have relatively low SNR. 

\subsubsection{TDE 2023ant}

2023ant was first reported as a ZTF alert on 2023-01-13 and was observed by DESI on 2023-02-27. We classify 2023ant as a TDE-H-He. We give this classification due to the broad He and H$\alpha$ emission present in the spectrum as well as the blackbody shape. We show a spectrum taken 43 days after discovery and 20 days post peak. The object's light curve shows a slow rise and decay and minimal color evolution ($g-r < 0.2$; Figure \ref{fig:tde_data}). 

\begin{figure*}
    \centering
    \includegraphics[width=0.49\textwidth]{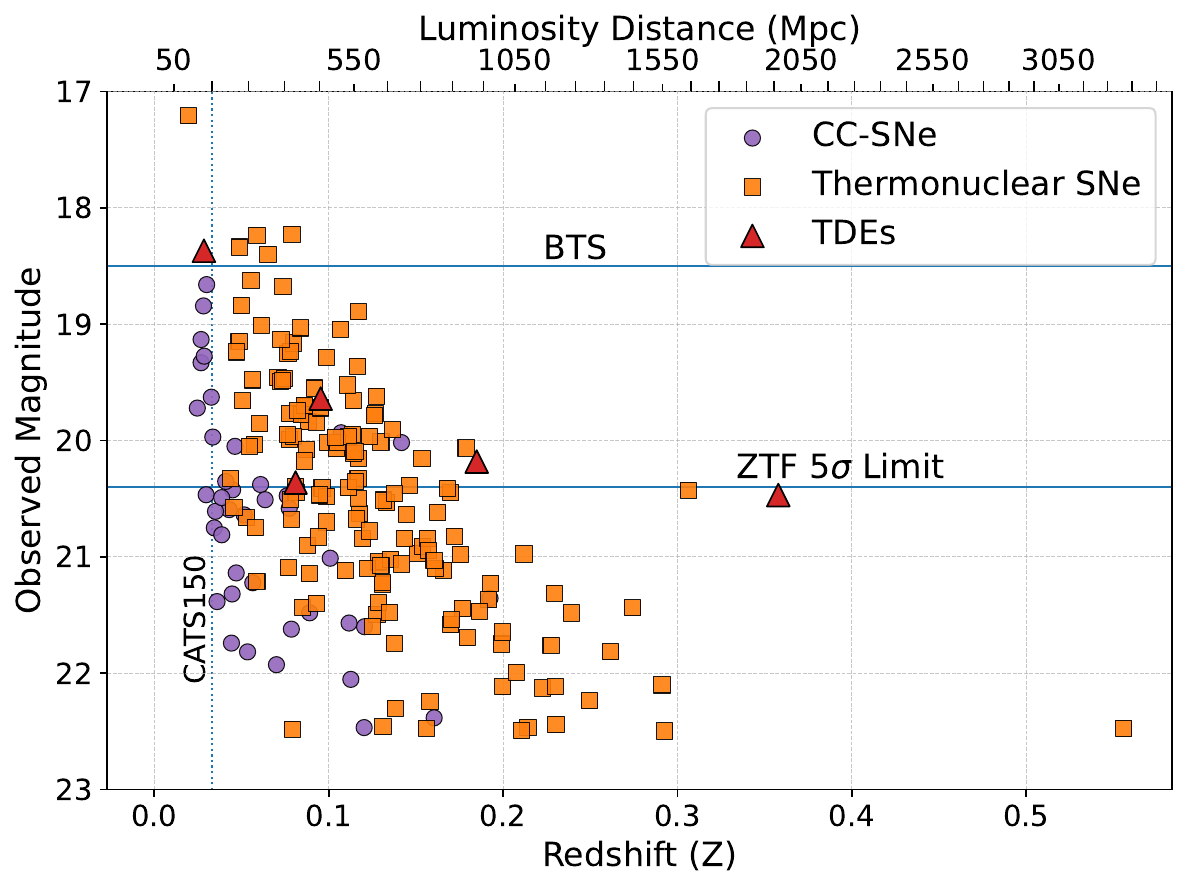}
    \includegraphics[width=0.49\textwidth]{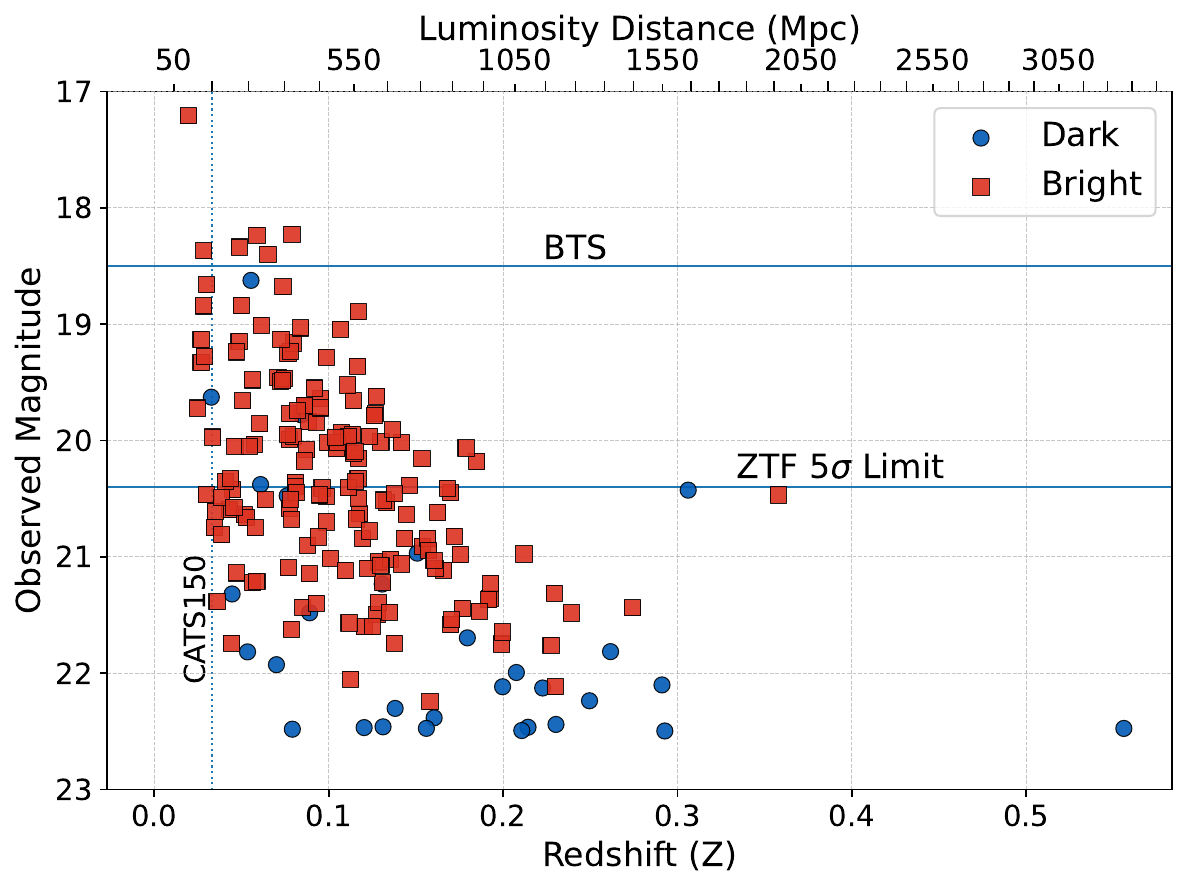}
    \caption{Computed LSST r-band AB magnitude from the DESI spectra plotted against their respective redshifts. Note almost half of the transients were observed when they were fainter than the $5\sigma$ cutoff for ZTF alerts. The lines drawn on the figure are the BTS spectroscopic cutoff of 18.5 magnitudes, the distance limit for follow-up imposed by CATS-150 (150 Mpc; \citealt{das_ztf_2025}), and the median single visit $5\sigma$ limit for a ZTF alert ($r_m \approx 20.4$; \citealt{dekany_zwicky_2020}). This demonstrates the immense power of DESI to follow-up transients down to a magnitude fainter than 22. \emph{Left:} transients color-coded by their class. \emph{Right:} transients color-coded by the DESI program they were observed under. It is clear that classification of mag$>22$ transients were mostly possible with the ``dark'' DESI program.}
    \label{fig:redshift_vs_obsmag}
\end{figure*}

\subsection{Observed Magnitudes}

The distribution of apparent magnitudes for DESI serendipitous and spare‐fiber transients extends significantly fainter than in previous systematic spectroscopic follow‐up programs. Surveys such as the BTS imposed hard cuts ($m < 18.5$ mag) that limited classifications to relatively nearby bright events. In contrast, the DESI spare‐fiber program routinely targets transients as faint as $m \lesssim 22$ mag for ToO candidates. The transient depth reached can be seen in Figure \ref{fig:redshift_vs_obsmag}. 
It is clear that even transients with mag $22-22.5$ can be classified with the DESI dark time program. The bright tiles have effective exposure times of 180 s versus the 1000 s of dark time. It is also remarkable that transients down to mag 22 are classified with the bright program. This depth makes DESI an ideal instrument to systematically observe transients discovered by LSST ( $r< 24.5$ mag) and DESIRT ( $r< 23.5$ mag).
By filling otherwise unused fibers with $m \lesssim 22$ mag transients, the spare‐fiber program adds $\sim$5–20 new classifications per night without impacting the primary galaxy survey. This efficient use of spare fibers not only boosts the total number of high‐redshift transients in DESI’s sample but also strengthens the statistical power for cosmological and progenitor studies at intermediate distances. 

\begin{figure}
    \centering
    \includegraphics[width=1\linewidth]{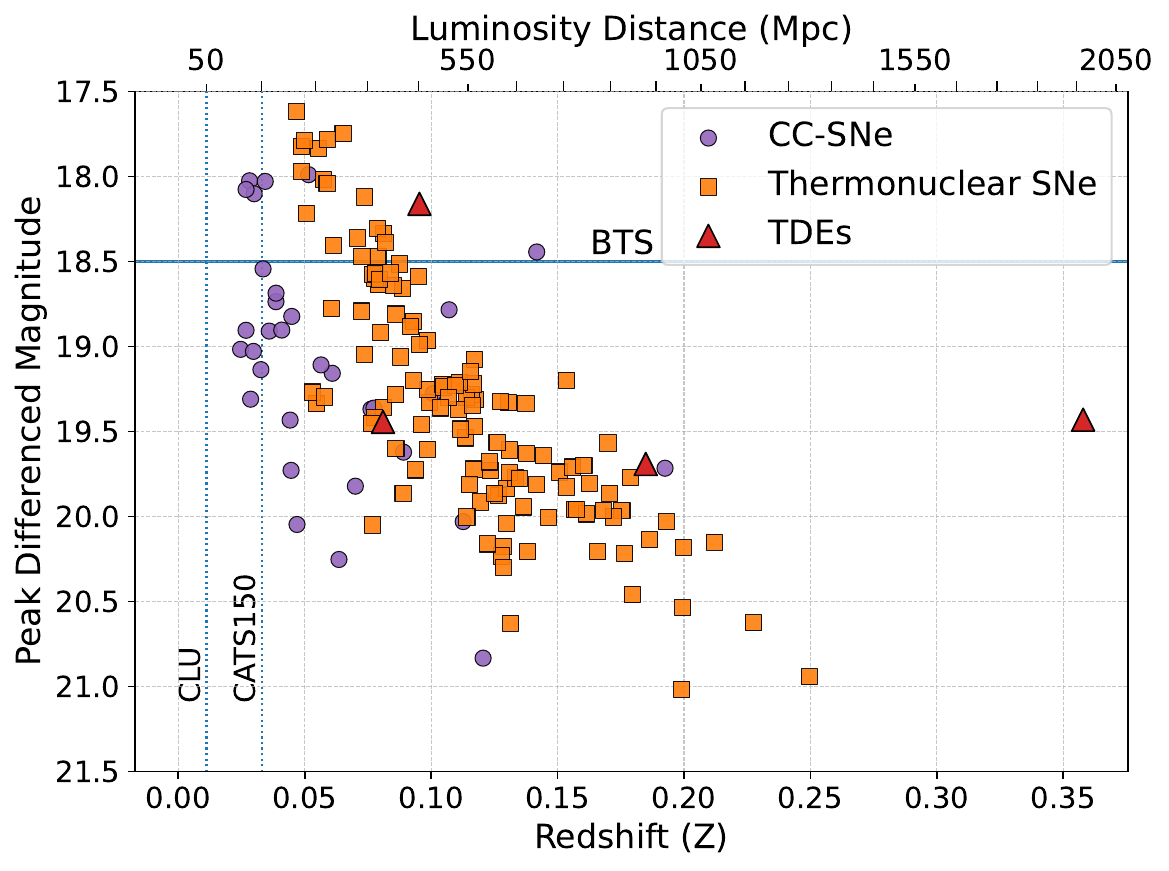}
    \caption{Observed peak magnitude in any band of the ZTF forced difference photometry versus the spectroscopically observed redshift. It can be noted that the majority of the sample exists below the cutoff enforced by the BTS. This spare fiber survey offers the ability to provide better redshift completeness for ZTF, DECam, and LSST targets at fainter magnitudes.}
    \label{fig:redshift_vs_peakmag}
\end{figure}

\subsection{Redshift Distribution}

As the sample of transients considered here was mainly discovered by ZTF, that limits the potential peak magnitude of most of the objects, as well as their number density. The redshift distribution of the DESI transients sample ranges from $0.05 < z < 0.25$ (Figure \ref{fig:redshift_vs_peakmag}), which covers an important gap in cosmological distributions. Many large samples cover either high-$z$ or low-$z$ type Ia SNe leaving little discovery for SNe caught between. This creates an uncomfortable gap in the Hubble diagram \citep{des_collaboration_dark_2024}. Furthermore, with the advent of LSST and cadence provided by DESIRT, it will be possible to build a population of spectrally classified thermonuclear SNe with well-defined light curves. Pushing $\sim$3 mag deeper increases the median redshift of the classified sample: Type Ia SNe can be followed  up to $z \simeq 0.6–0.7$, and core collapse SNe to $z \simeq 0.3$, roughly doubling the redshift reach of current spectroscopic efforts. Thus, the ongoing DESI spare fiber program once expanded to DESIRT and LSST discovered thermonuclear SNe will be able to bridge a necessary gap and better constrain the systemic errors that plague better constraints on the Hubble parameter.

\begin{figure*}
    \includegraphics[width=0.98\textwidth]{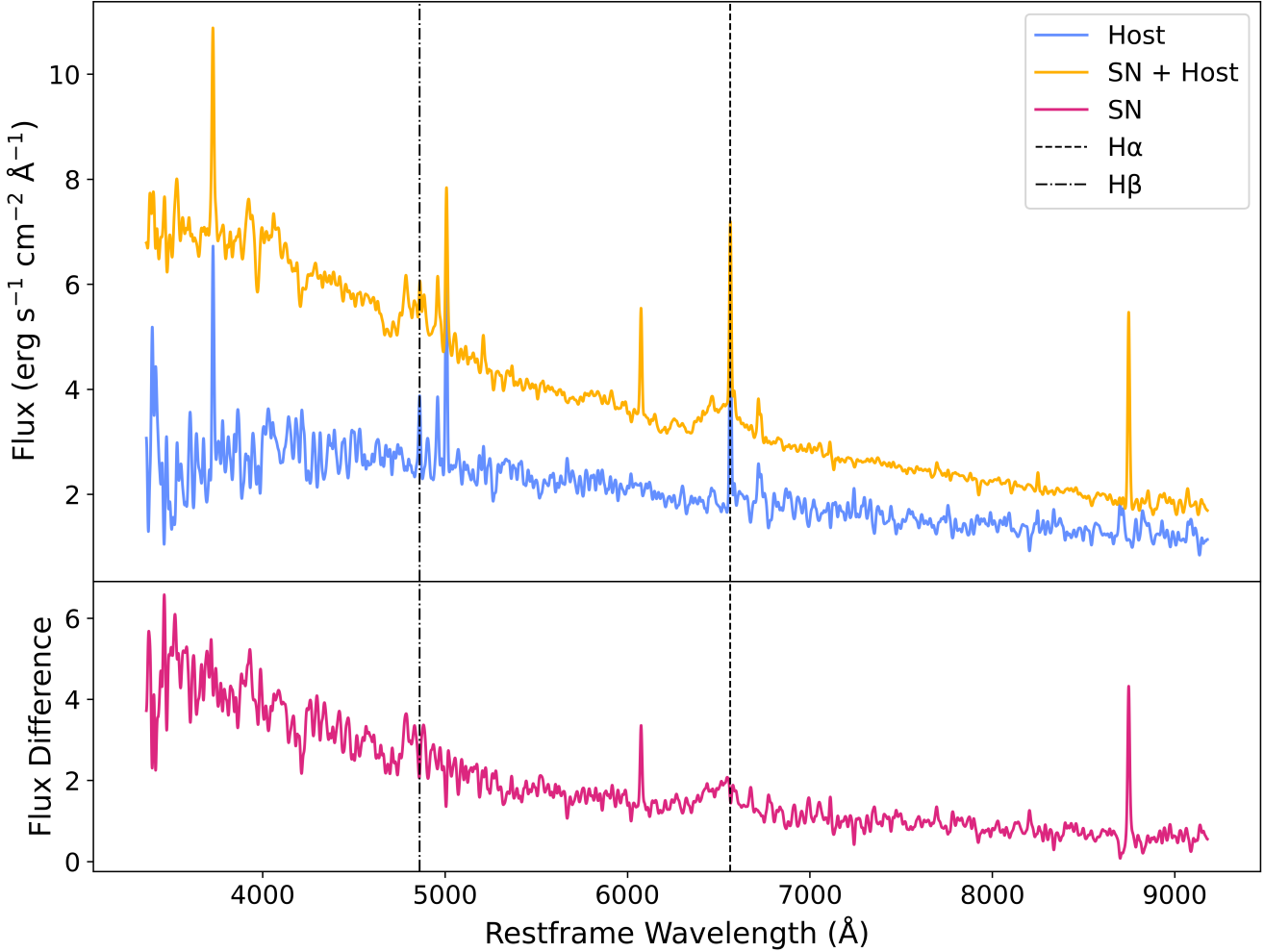}
    \caption{The Host spectrum and the unsubtracted flare spectrum of 2025iip are on the top. Notice the strong galaxy lines that make the underlying broad SNe features more difficult to resolve. However, in the lower panel the subtracted spectrum shows clear broad Hydrogen features with most of the contaminating galaxy lines removed.}
    \label{fig:2025iip}
\end{figure*}

\subsection{Host Subtraction}

A major advantage of the DESI spare fiber survey is the capability to produce host-subtracted transient spectra. DESI has already observed over 40 million galaxies so it is inevitable that new transients will be discovered in these hosts. Then this spare fiber program can re-observe the novel transient. Such a situation avoids many of the statistical and physical complexities of attempting to subtract hosts from different instruments, with different apertures and resolutions \citep[see][for a method of using DESI hosts in this case]{hall_at2025ulz_2025}. These ideal host subtractions can help classify transients that would otherwise be misclassified. Here we present the situation of 2025iip and the potential power of DESI's ability to host subtract its own transients.


AT 2025iip was discovered on 2025-04-20 by ZTF, reported to TNS by the ALeRCE group on 2025-04-24 \citep{carrasco-davis_alert_2021}, and submitted to the DESI spare fiber program on 2025-04-25. It was observed by DESI on 2025-04-30, appearing as a blue spectrum with many galaxy lines. These galaxy lines overwhelmed the small H$\alpha$ bump that was present in the spectrum making it difficult to tell if the bump was simply from galaxy lines or not. However, with the host subtracted it became clear that 2025iip had both broad H$\alpha$ and H$\beta$ lines with potential P Cygni distributions (Figure \ref{fig:2025iip}). After further monitoring of its light curve it slowly began to redden which confirmed its classification as a Type II SN. Host subtraction such as this is essential for accurate emission line analysis of SNe and TDEs, as without it, a transient's observed spectra are a composition of both their underlying host and the transient itself \citep{charalampopoulos_detailed_2022}. 

\section{Conclusions}\label{sec:conclusions}

 We show results from the DESI Transients Survey, a DESI spare fiber program which is routinely observing a range of extragalactic astronomical transients. One of the main outcomes of the DESI Transient Survey will be its ability to keep up with the Rubin LSST transient discovery density, at least down to mag $\sim 22$, down to which we have shown to be able to systematically classify publicly reported transients. DESIRT is a  pathfinder for the difficulties in target count that will be experienced once LSST alerts start to be issued in 2026. The spike in number of ToOs observed by DESI and caused by the restart of DESIRT in 2025 is an important signal for the number of transients that will be achieved by the regular observations of the Rubin LSST wide fast deep survey. Indeed, DESI observations of DESIRT targets reached over 10 on some of the best nights, but covered at least an order of magnitude less area than what Rubin will do, thus numbers could significantly rise during LSST operations. As expected since the target density is still quite sparse ($\sim 1$ per deg$^2$ per day), the percent completion per night of transients remained constant both before and after the restart of the DESIRT program ($\sim 13\%$). Thus, with the even greater number of transients discovered by LSST, this DESI spare fiber program will be ready to provide  $\mathcal{O}(1000)$ valuable transient spectra per year. 

As part of the early phase of this program, we release the classification and spectra for over 250 transients including tidal disruption events, thermonuclear and core-collapse SNe. An interesting result of our findings is the discovery and classification of a young TDE. Through the serendipitous observations, DESI was able to observe this TDE seven days before the transient’s discovery by photometric surveys, thus securing valuable early-time spectral observations \citep{andreoni_very_2022}. This unprecedented catch underscores the unique capability of DESI to identify and study ``infant” transients, providing new insight into the early evolution of tidal disruption phenomena. With the new spare fiber program and the upcoming Rubin observations, such rare events will become more common with more infant TDEs and SNe being observed with high quality DESI spectra. 

In parallel, the classifications from DESI have assembled a substantial sample of previously unknown Type Ia supernovae at intermediate redshifts ($0.05 < z < 0.25$). This sample fills a key gap between the traditional low-$z$ and high-$z$ SN~Ia populations, thereby enhancing the statistical power of cosmological analyses \citep{des_collaboration_dark_2024}. By observing SNe~Ia in this intermediate range, the DESI Transient Survey can help mitigate systematic uncertainties in Hubble parameter measurements.

Furthermore, CCSNe at these redshifts will offer insight into the chemical evolution of galaxies, enable new rate estimates, and also supernova cosmology measurements \citep{melinder_rate_2012, boccioli_physics_2024}. Finally, TDEs will allow for the study of dormant massive black holes, their mass function, and are essential to study stellar populations and dynamics in galactic nuclei at intermediate redshifts \citep{graur_dependence_2018, velzen_seventeen_2021, hammerstein_final_2022, yao_tidal_2023, mockler_weighing_2019, mockler_uncovering_2023}. The higher resolution of the DESI spectrograph, compared to other instruments such as the Spectral Energy Distribution Machine (SEDM; \citealt{blagorodnova_sed_2018}, will also allow for better identification of unique transients, such as those with Fe coronal lines \citep{short_delayed_2023}, at these intermediate redshifts.

Together, these discoveries highlight the promise of the DESI ToO spare fiber program and its potential for the future. Using otherwise idle fibers, this program efficiently increases the yield of transient classifications each night without detracting from the primary galaxy survey. Given current rates with the DESI ToO program, we will expect to discover on order of ${\sim}500$ SN a year at these intermediate redshifts. Looking ahead, the synergy between DESI’s rapid spectroscopic follow-up and deep, high-cadence imaging surveys such as DESIRT and the Rubin Observatory’s LSST will be particularly powerful.

The next generation of imaging surveys will vastly expand the discovery space of faint and fast-evolving transients, and the DESI spare-fiber strategy can promptly provide spectroscopic confirmation and characterization for 50\% of these targets within 8 days of triggering (Figure \ref{fig:timeframe}). For example, reaching ${\sim}3$ magnitudes deeper than current transient searches would roughly double the redshift horizon of spectroscopically classified SNe~Ia (to $z \sim 0.6–0.7$), while also extending core-collapse SN follow-up to $z \sim 0.3$. The continued spare-fiber program, augmented by upcoming surveys like DESIRT and LSST stands poised to routinely capture early-stage phenomena and to build statistically rich transient samples at intermediate and higher redshifts. Such capabilities pave the way for new insights into transient astrophysics and offer improved constraints on cosmology in the LSST era.

\section{Data Availability and Releases}


The classifications obtained from the DESI spectra presented in this work are released to the community. At the time of publication of this work, all observed spectra, along with their respective classifications, from DESI DR1 are accessible on TNS. Furthermore, as the DESI spare-fiber program continues, new classifications from DESI spectra will be reported to TNS with future releases. The spectra identified as related to transients within this paper, as well as machine readable files with classifications are provided at \citet{hall_dts_zenodo_2025}.

\begin{acknowledgments}

AP is supported by NSF Grant No. 2308193. B.O. acknowledges support from the McWilliams Fellowship at Carnegie Mellon University.

This material is based upon work supported by the U.S. Department of Energy (DOE), Office of Science, Office of High-Energy Physics, under Contract No. DE–AC02–05CH11231, and by the National Energy Research Scientific Computing Center, a DOE Office of Science User Facility under the same contract. Additional support for DESI was provided by the U.S. National Science Foundation (NSF), Division of Astronomical Sciences under Contract No. AST-0950945 to the NSF’s National Optical-Infrared Astronomy Research Laboratory; the Science and Technology Facilities Council of the United Kingdom; the Gordon and Betty Moore Foundation; the Heising-Simons Foundation; the French Alternative Energies and Atomic Energy Commission (CEA); the National Council of Humanities, Science and Technology of Mexico (CONAHCYT); the Ministry of Science, Innovation and Universities of Spain (MICIU/AEI/10.13039/501100011033), and by the DESI Member Institutions: \url{https://www.desi.lbl.gov/collaborating-institutions}.

The DESI Legacy Imaging Surveys consist of three individual and complementary projects: the Dark Energy Camera Legacy Survey (DECaLS), the Beijing-Arizona Sky Survey (BASS), and the Mayall z-band Legacy Survey (MzLS). DECaLS, BASS and MzLS together include data obtained, respectively, at the Blanco telescope, Cerro Tololo Inter-American Observatory, NSF’s NOIRLab; the Bok telescope, Steward Observatory, University of Arizona; and the Mayall telescope, Kitt Peak National Observatory, NOIRLab. NOIRLab is operated by the Association of Universities for Research in Astronomy (AURA) under a cooperative agreement with the National Science Foundation. Pipeline processing and analyses of the data were supported by NOIRLab and the Lawrence Berkeley National Laboratory. Legacy Surveys also uses data products from the Near-Earth Object Wide-field Infrared Survey Explorer (NEOWISE), a project of the Jet Propulsion Laboratory/California Institute of Technology, funded by the National Aeronautics and Space Administration. Legacy Surveys was supported by: the Director, Office of Science, Office of High Energy Physics of the U.S. Department of Energy; the National Energy Research Scientific Computing Center, a DOE Office of Science User Facility; the U.S. National Science Foundation, Division of Astronomical Sciences; the National Astronomical Observatories of China, the Chinese Academy of Sciences and the Chinese National Natural Science Foundation. LBNL is managed by the Regents of the University of California under contract to the U.S. Department of Energy. The complete acknowledgments can be found at \url{https://www.legacysurvey.org/}.

Any opinions, findings, and conclusions or recommendations expressed in this material are those of the author(s) and do not necessarily reflect the views of the U. S. National Science Foundation, the U. S. Department of Energy, or any of the listed funding agencies.

The authors are honored to be permitted to conduct scientific research on Iolkam Du’ag (Kitt Peak), a mountain with particular significance to the Tohono O’odham Nation.

The ZTF forced-photometry service was funded under the Heising-Simons Foundation grant \#12540303 (PI: Graham).
\end{acknowledgments}
%
\facilities{CTIO:Blanco, KPNO:Mayall.}

\software{Astropy \citep{2013A&A...558A..33A, astropy_collaboration_astropy_2018, astropy_collaboration_astropy_2022}, Numpy \citep{harris2020array}, 
Redrock \citep{Redrock.Bailey.2024},
          NGSF \citep{goldwasser_next_2022}.
          }


\appendix



\section{SNe Sample}\label{app}

Below is a spectroscopic randomly chosen sample of 8 Type Ia SNe, 8 Type II SNe, and 5 Type Ib/c SNe that were classified by this program. We also include our time dependent sample of spectra from SN 2023ixf, one of the closest type II SN to date \citep{kilpatrick_sn_2023}.

\begin{figure*}
    \centering
    \includegraphics[width=0.98\textwidth]{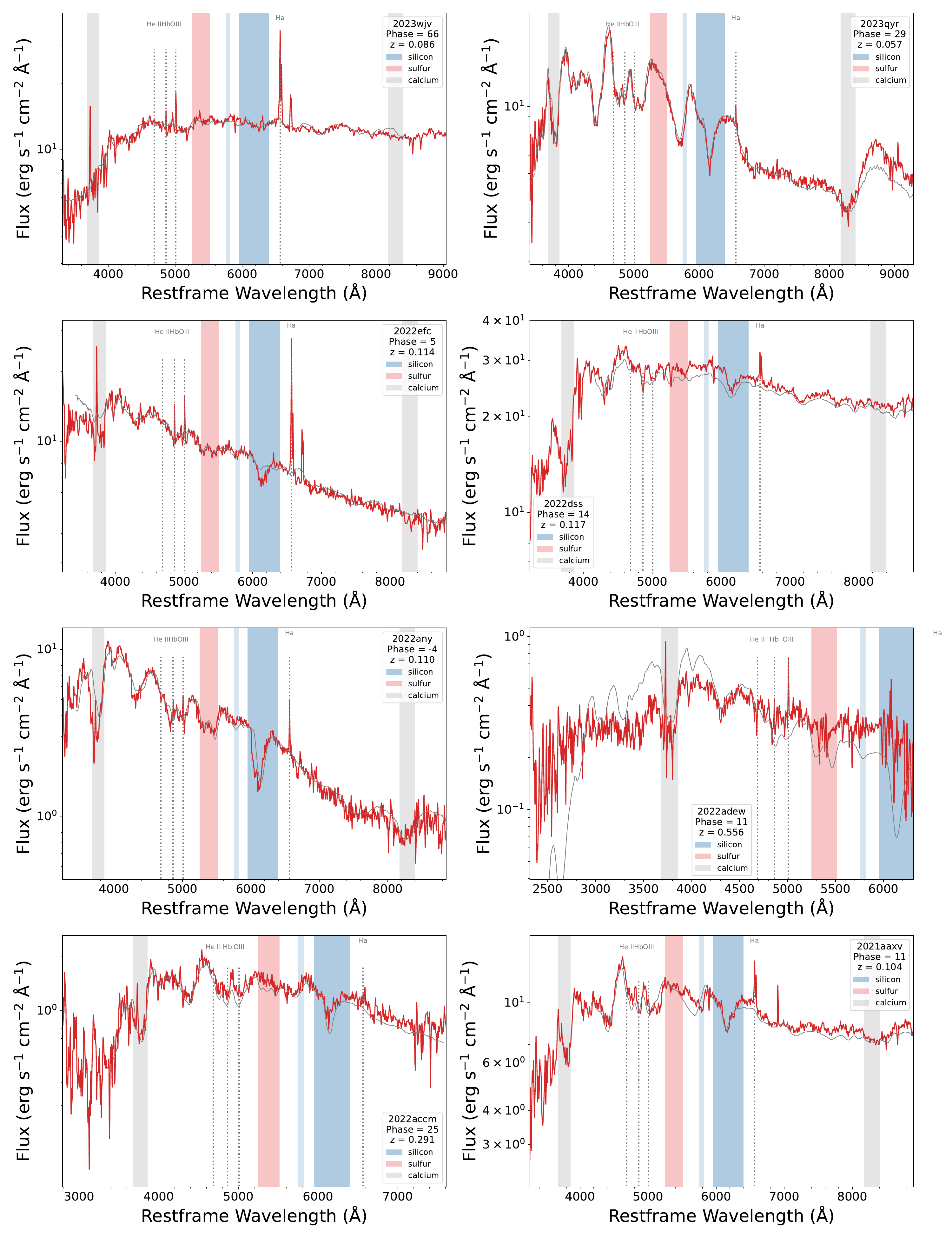}
    \caption{Spectroscopic sample of classified type Ia SNe in DESI in their restframe wavelength with a log scaled y-axis. Specific galaxy lines are noted in dotted lines while broad SNe features are highlighted in their respective color. The ``best fit" template from Superfit is plotted in gray.}
    \label{fig:Ia_spectra}
\end{figure*}

\begin{figure*}
    \centering
    \includegraphics[width=0.98\textwidth]{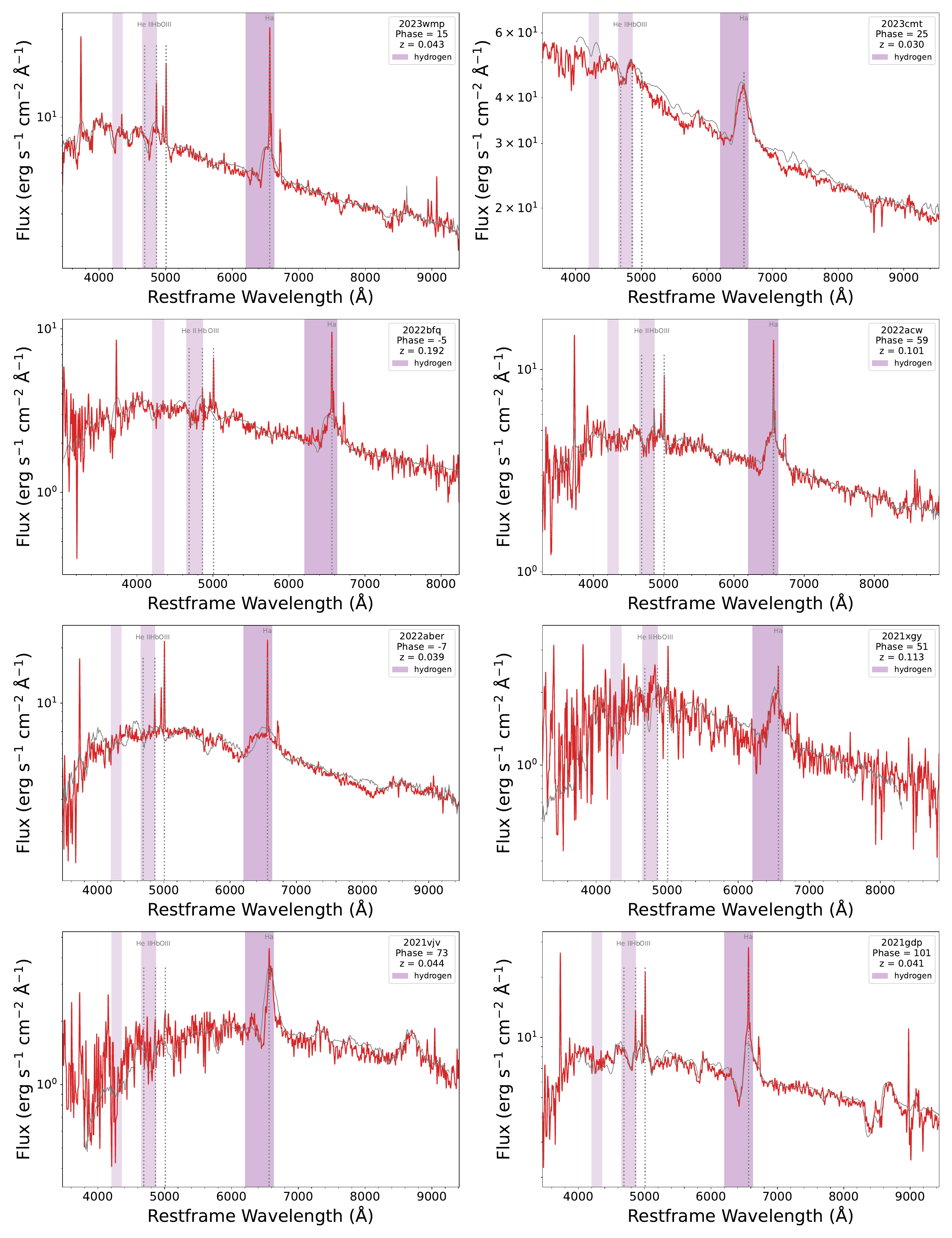}
    \caption{Spectroscopic sample of classified type II SNe in DESI in their restframe wavelength with a log scaled y-axis. Specific galaxy lines are noted in dotted lines while broad SNe features are highlighted in their respective color. The ``best fit" template from Superfit is plotted in gray.}
    \label{fig:II_spectra}
\end{figure*}

\begin{figure*}
    \centering
    \includegraphics[width=0.98\textwidth]{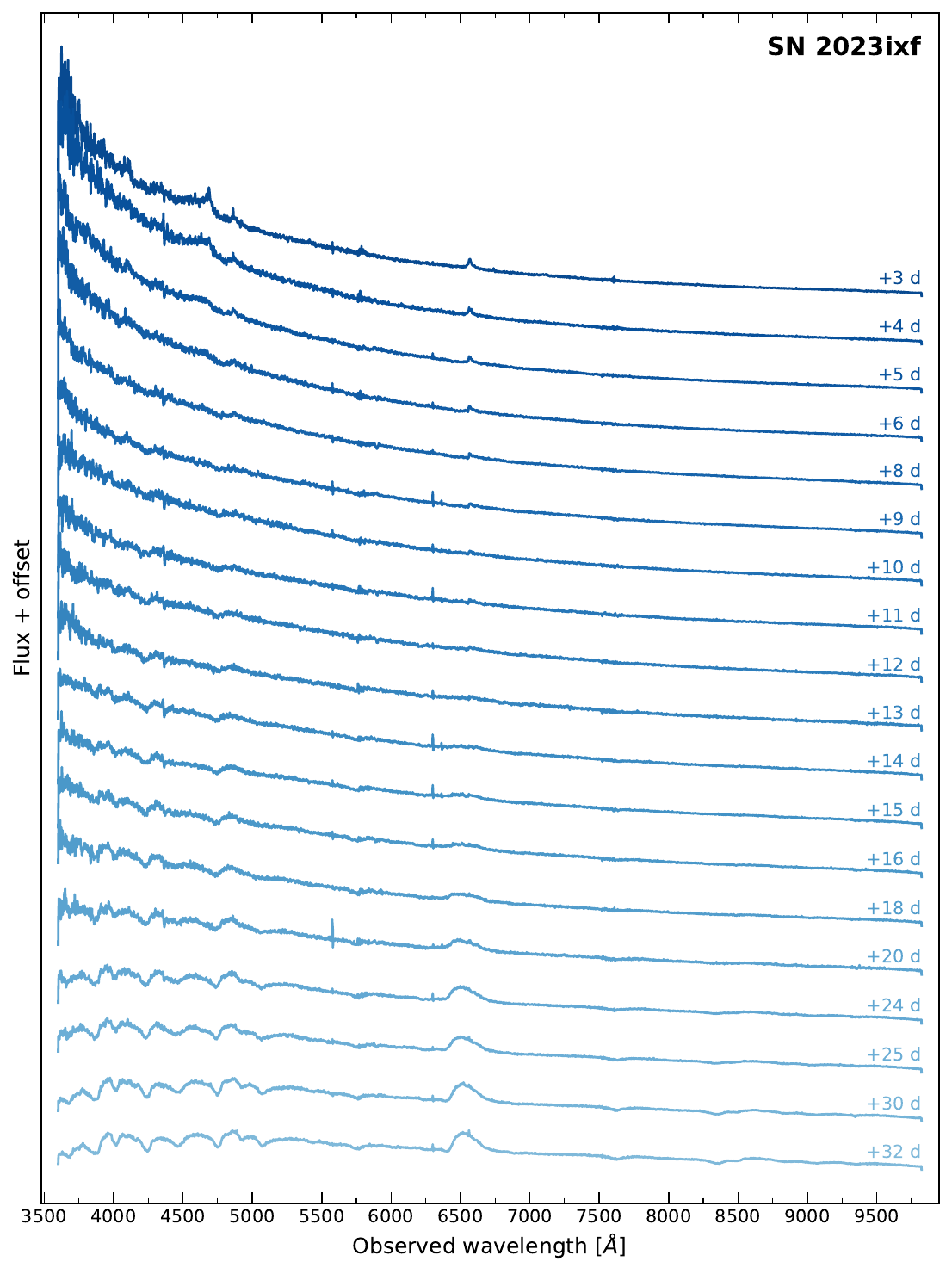}
    \caption{Spectroscopic series of the type II SN 2023ixf. During twilight time, DESI was able to observe this transient 19 times in a month. The time is given since the discovery time of SN 2023ixf.}
    \label{fig:SN2023ixf}
\end{figure*}

\begin{figure*}
    \centering
    \includegraphics[width=0.98\textwidth]{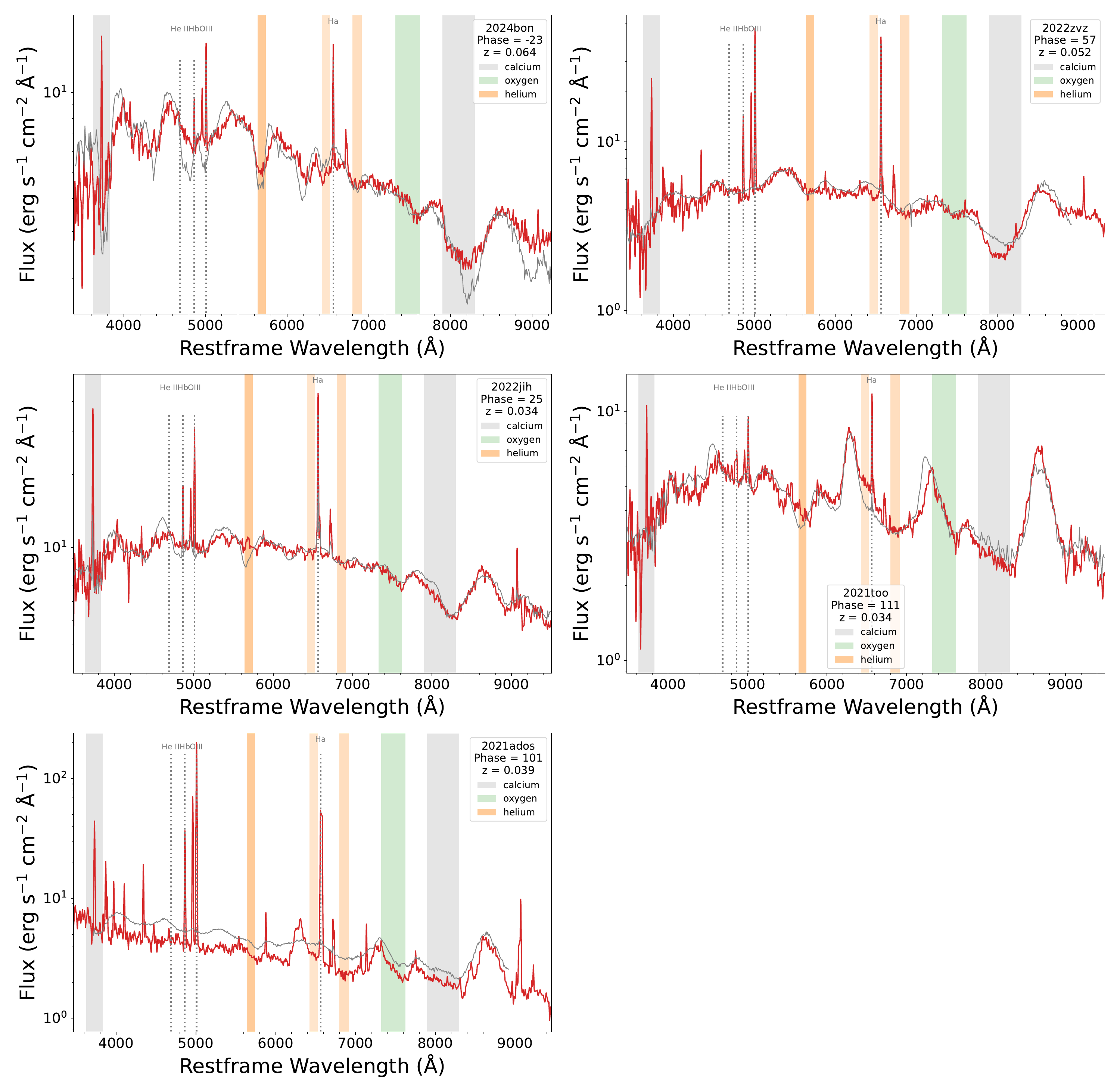}
    \caption{Spectroscopic sample of classified type Ib/c SNe in DESI in their restframe wavelength with a log scaled y-axis. Specific galaxy lines are noted in dotted lines while broad SNe features are highlighted in their respective color. The ``best fit" template from Superfit is plotted in gray.}
    \label{fig:Ibc_spectra}
\end{figure*}


\bibliography{TDESearch,DESITransientPaper,sample7}{}
\bibliographystyle{aasjournal}



\end{document}